\documentclass[fleqn,usenatbib]{mnras}
\usepackage{hyperref}
\usepackage[T1]{fontenc}
\usepackage{ae,aecompl}
\usepackage{graphicx}
\usepackage{amsmath}
\let\oldAA\AA
\renewcommand{\AA}{\text{\normalfont\oldAA}}
\usepackage{array}
\usepackage{makecell}
\usepackage{amssymb}
\usepackage{booktabs}
\usepackage{pdflscape} 
\usepackage{footnote}
\usepackage{threeparttable}
\usepackage{epsfig}
\usepackage{pslatex}
\usepackage{csquotes}
\usepackage{multirow}
\usepackage{orcidlink}
\usepackage{soul}
\usepackage{isotope}
\usepackage{pdflscape}
\graphicspath{{./}{figures/}}
\title[V2891 Cyg]{A phenomenological study of the evolution of shock-induced \ion{O}{i} emission lines in the spectrum of nova V2891 Cygni}
\author[]{
Ruchi Pandey\orcidlink{0000-0002-6222-3045},$^{1}$\thanks{E-mail:ruchi@prl.res.in, ruchie20pandey@gmail.com}
Mudit K. Srivastava\orcidlink{0000-0002-0138-7601}, $^{1}$\thanks{Email:mudit@prl.res.in}
Gargi Shaw\orcidlink{0000-0003-4615-8009} $^{2}$\\
$^{1}$ Astronomy \& Astrophysics Division, Physical Research Laboratory, Ahmedabad 380009, Gujarat, India\\
$^{2}$ Department of Astronomy and Astrophysics, Tata Institute of Fundamental Research, Homi Bhabha Road, Mumbai 4000005, India\\ 
}

\begin{document}
\label{firstpage}
\pagerange{\pageref{firstpage}--\pageref{lastpage}}
\maketitle
\begin{abstract}
The eruption of Nova V2891 Cygni in 2019 offers a rare opportunity to explore the shock-induced processes in novae ejecta. The spectral evolution shows noticeable differences in the evolution of various oxygen emission lines such as \ion{O}{i} 7773 {\AA}, \ion{O}{i} 8446 {\AA}, \ion{O}{i} 1.1286 $\mu$m, \ion{O}{i} 1.3164 $\mu$m, etc. Here, we use spectral synthesis code \textsc{cloudy} to study the temporal evolution of these oxygen emission lines. Our photoionization model requires the introduction of a component with a very high density ($n \sim 10^{11}$ cm$^{-3}$) and an enhanced oxygen abundance (O/H $\sim$ 28) to produce the \ion{O}{i} 7773 {\AA} emission line, suggesting a stratification of material with high oxygen abundance within the ejecta. An important outcome is the behaviour of the \ion{O}{i} 1.3164 $\mu$m line, which could only be generated by invoking the collisional ionization models in \textsc{cloudy}.  Our phenomenological analysis suggests that \ion{O}{i} 1.3164 $\mu$m emission originates from a thin, dense shell characterized by a high density of about $10^{12.5} - 10^{12.8}$ cm$^{-3}$, which is most likely formed due to the strong internal collisions. If such is the case, the \ion{O}{i} 1.3164 $\mu$m emission presents itself as a tracer of shock-induced dust formation in V2891 Cyg. The collisional ionization models have also been successful in creating the high-temperature conditions ($\sim 7.07 - 7.49 \times 10^5$ K) required to reproduce the observed high ionization potential coronal lines, which coincide with the epoch of dust formation and evolution of the \ion{O}{i} 1.3164 $\mu$m emission line.
\end{abstract}
\begin{keywords} 
stars: individual (V2891 Cyg) - novae, cataclysmic variables
\end{keywords}

\section{Introduction}\label{sec1}

Classical novae (hereafter, CNe) outbursts are cataclysmic events characterized by an abrupt increase in the brightness of a star. CNe occur in a binary star system where a primary white dwarf (hereafter, WD) accretes hydrogen-rich material from its companion star. The primary WD is usually a carbon-oxygen (CO) with relatively low mass ($M_{WD} < 1.2 M_{\odot}$) or oxygen-neon (ONe) type with relatively high mass ($M_{WD} > 1.2 M_{\odot}$), whereas the secondary is usually a main-sequence star \citep[][and references therein]{2024MNRASEvans}. When a sufficient mass has accumulated, the conditions for thermonuclear reactions are satisfied, resulting in the sudden ignition of hydrogen fusion. The ignition results in a thermonuclear runaway, which ejects a cloud of gas and other material into space. Around $\sim 10^{-4} - 10^{-5}$ M${\odot}$ of this layer is ejected from the system in each outburst \citep[see e.g.][for a detailed review]{1998PASP..110....3G,2008clno.book.....B,2020ApJStarrfield}.
\par
The classical nova V2891 Cyg (RA: 21:09:25.52, Dec: +48:10:51.9), which is also referred to as AT 2019qwf, PGIR 19brv, and ZTF19abyukuy, was detected on UT 2019-09-17.25 (JD 245 8743.75) by \citet{2019ATel13130....1D}. The recorded magnitude in the J band was 11.3 $\pm$ 0.02, as reported by \citet{2019ATel13130....1D}. \citet{2022MNRAS.510.4265K} suggested that the initial detection of the nova occurred on 2019 September 14.16 UT (JD 245 8740.66) with a magnitude of r = 19.34 $\pm$ 0.18 during the early stages of its rise from the quiescence phase. We consider this date ($t_0 = $ JD 245 8740.66) as the date of the outburst in this study. V2891 Cyg is a highly reddened galactic nova \citep{2019ATel13130....1D}. The optical and infrared studies classified V2891 Cyg as an \ion{Fe}{ii} type nova \citep[][and references therein]{2019ATel13283....1M,2019ATel13149....1L,2019ATel13258....1S,2019ATel13301....1J}. The rate of decline for the V2891 Cyg was estimated to be t$_2 \sim$ 100-150 days, and t$_3 \sim $ 180-230 days, where t$_2$ and t$_3$ represent the number of days taken by the nova to decrease in brightness by 2 and 3 magnitudes, respectively, from its peak brightness, making it a slow nova \citep{2022MNRAS.510.4265K}. The reddening and distance to the nova are estimated to be $E(B - V) = 2.21 \pm 0.15$ and d = 5.50 kpc, respectively \citep{2022MNRAS.510.4265K}. Recently, utilising the latest Gaia data release, \citet{2022MNRASSchaefer} estimated the distance and reddening to be d = 6991$^{+6761}_{-2141}$ pc and $E(B - V) = 2.70 \pm 0.90$, respectively. Their reddening estimation surpasses the value of reddening estimated by \citet{2022MNRAS.510.4265K}. However, we notice that the reddening estimated by \citet{2022MNRAS.510.4265K} is within 1$\sigma$ of the value reported by \citet{2022MNRASSchaefer}. As in this work, we have attempted to explain the observed behaviour of V2891 Cyg by \citet{2022MNRAS.510.4265K} through a phenomenological approach, we have opted to use the values derived from their study to deredden the spectra. V2891 Cyg showed the evolution of various oxygen emission lines from the initial fireball phase to the coronal phase over 15 months following the outburst \citep{2022MNRAS.510.4265K}. For a detailed discussion on the spectral evolution, please see \citet{2022MNRAS.510.4265K}. V2891 Cyg showed two peculiar behaviours during its evolution. One of these features involved the presence of a time-varying P-Cygni profile in the emission line of \ion{O}{i} at 7773 {\AA}. This behaviour may imply the potential occurrence of multiple distinct episodes of mass ejection within the ejecta \citep[][and references therein]{2020ApJAydi}. Furthermore, the nova had a short period of dust formation that occurred simultaneously with the phase of coronal line emission. Based on these observations, \citet{2022MNRAS.510.4265K} proposed the existence of multiple outflows of different velocities in the ejecta, as indicated by the time-varying P-Cygni profile in the emission line of \ion{O}{i} at 7773 {\AA}. The collision of discrete mass outflows resulted in internal shocks within the ejecta, which could have led to the dust formation, as suggested in recent literature \citep{Derdzinski2017MNRAS}. In their study, \citet{2022MNRAS.510.4265K} provided evidence supporting the hypothesis that the origin of coronal lines is attributed to shock heating rather than photoionization. Considering all these factors, \citet{2022MNRAS.510.4265K} proposed the hypothesis that shock-induced processes are accountable for dust formation in V2891 Cyg.
\par
Observations have revealed that a significant number of novae generate dust and molecules following an eruption. Despite the harsh conditions found in novae, dust forms relatively quickly after an outburst, typically within mere months \citep{2008clno.book.....B}. However, the dust formation process within the expanding nova ejecta has remained an enigma for decades. Numerous studies have been done to elucidate the dust formation process in novae. These studies focus on specific scenarios, such as the chemistry of nova ejecta \citep{1988MNRAS507Rawlings,1989MNRASRawlings,2004MNRASPontefract}, CO emission associated with dusty novae \citep{2003ApJRudy}, CNO enhancement in the ejecta \citep{1988ARA&AGehrz}, kinetic agglomeration and photoionization processing \citep{Shore2004A&A}, etc. Despite numerous investigations spanning several decades, the mechanisms underpinning the formation of dust grains in the nova ejecta have remained elusive. Recently, a few studies have put forth the concept of internal shocks occurring within the ejecta of novae \citep[for example][and references therein]{Metzger2014MNRAS,Chomiuk2014Natur}, which can create an ideal environment for dust formation in novae. The cooling efficiency of the gas behind these shock regions results in a subsequent increase in density. Consequently, this leads to the creation of cold, neutral, and dense clumps of gas. These clumps are effectively shielded from the intense radiation emitted by the WD. As a result, grain formation occurs rapidly within these protected regions, facilitating the efficient growth of grains to substantial sizes \citep{Derdzinski2017MNRAS}.
\par
In this study, we have generated a phenomenological model of V2891 Cyg to elucidate the observed optical and near-infrared (NIR) emission features and continuum. This work examines the evolution of prominent emission lines of oxygen, especially in conjunction with the time frame of dust formation, which also aligns with the phase of coronal emission observed in V2891 Cyg. Following the previously published studies on shock-induced dust formation by \citet{Derdzinski2017MNRAS}, and the observational study by \citet{2022MNRAS.510.4265K} we develop rudimentary phenomenological models utilising the photoionization code \textsc{cloudy}, v22.01 \citep[][and references therein]{2017RMxAA..53..385F}. The data analysed in this study are taken from \citet{2022MNRAS.510.4265K}. For a detailed discussion on observations and reductions, we refer the reader to their article.
\par
This paper is organised according to the following outline: Section~\ref{sec:oilines} discusses the temporal evolution of various oxygen lines and their potential excitation mechanisms observed in the spectra of V2891 Cyg. In particular, two of the \ion{O}{i} emission lines at 7773 {\AA} and 1.3164 $\mu$m show distinct evolutionary behaviour, and these aspects are also discussed in section~\ref{sec:oilines}. Section~\ref{sec:cloudy} presents a detailed analysis of the nebular phase in V2891 Cyg using \textsc{cloudy} photoionization models, which helps explain the observed behaviour of \ion{O}{i} 7773 {\AA} emission line. In Section~\ref{sec:OIlines_shock}, we explore the use of pure collisional ionisation models of \textsc{cloudy} to generate the \ion{O}{i} 1.3164 $\mu$m emission line and coronal emission lines observed in the spectra of the V2891 Cyg. The paper concludes with a summary and conclusion in Section~\ref{sec:summary}.

\section{Temporal evolution of O I emission lines}\label{sec:oilines}

The spectral evolution of V2891 Cyg, as presented by \citet{2022MNRAS.510.4265K}, in both optical and NIR bands, shows numerous prominent oxygen lines. During the evolution from the initial fireball phase to the coronal phase, the nova V2891 Cyg showed the most prominent oxygen lines of \ion{O}{i} 4959 {\AA}, [\ion{O}{iii}] 5007 {\AA}, [\ion{O}{i}] 5577 {\AA}, [\ion{O}{i}] 6300 {\AA}, [\ion{O}{ii}] 7320 {\AA} \ion{O}{i} 7773 {\AA}, and \ion{O}{i} 8446 {\AA} in the optical band and \ion{O}{i} 1.1286 {$\mu$m}, and \ion{O}{i} 1.3164 {$\mu$m} in the infrared domain. These \ion{O}{i} lines were detected at different stages of the evolution and showed variation in flux as the nova progressed through its outburst and decline phases. The appearance, disappearance, profile shape, line flux, etc., of these emission lines, offer a rare opportunity to study evolution of the ejecta over an extended period. In this study, we have used the air values of wavelength references in optical and IR bands. These wavelength reference values are identical to the ones used by \cite{2022MNRAS.510.4265K}.
\par
\begin{figure}
\centering
\includegraphics[height=8.5cm,width=9cm]{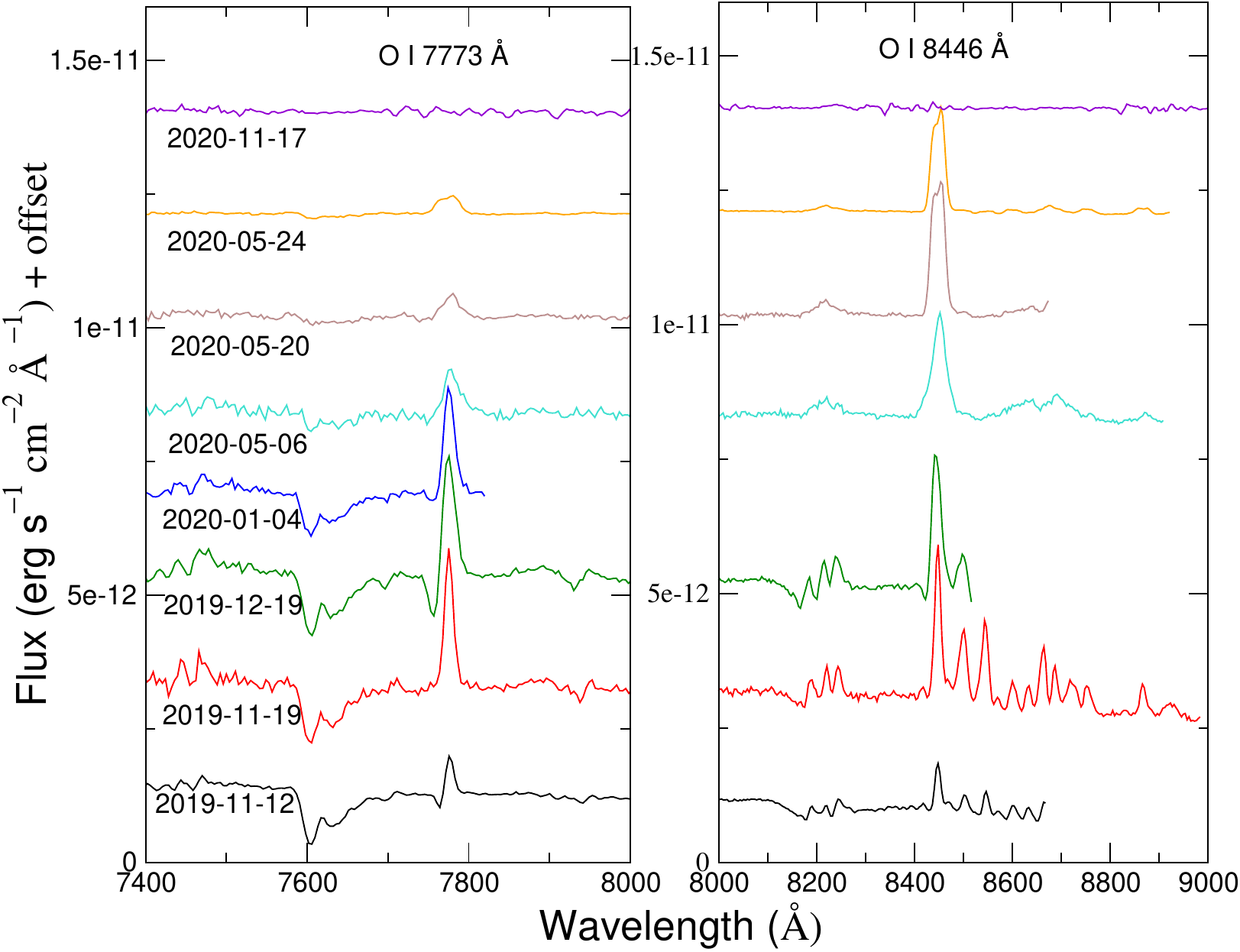}
\caption{Evolution of the prominent Oxygen emission lines \ion{O}{i} 7773, and 8446 {\AA} in the optical spectrum of nova V2891 Cyg. The spectra are reddening corrected. An offset of $2.0 \times 10^{-12}$ is added to the spectra. See Section~\ref{sec:oilines} for detailed discussion.}
\label{fig:optical}
\end{figure}
\begin{figure}
    \centering
    \includegraphics[height=8.5cm,width=9cm]{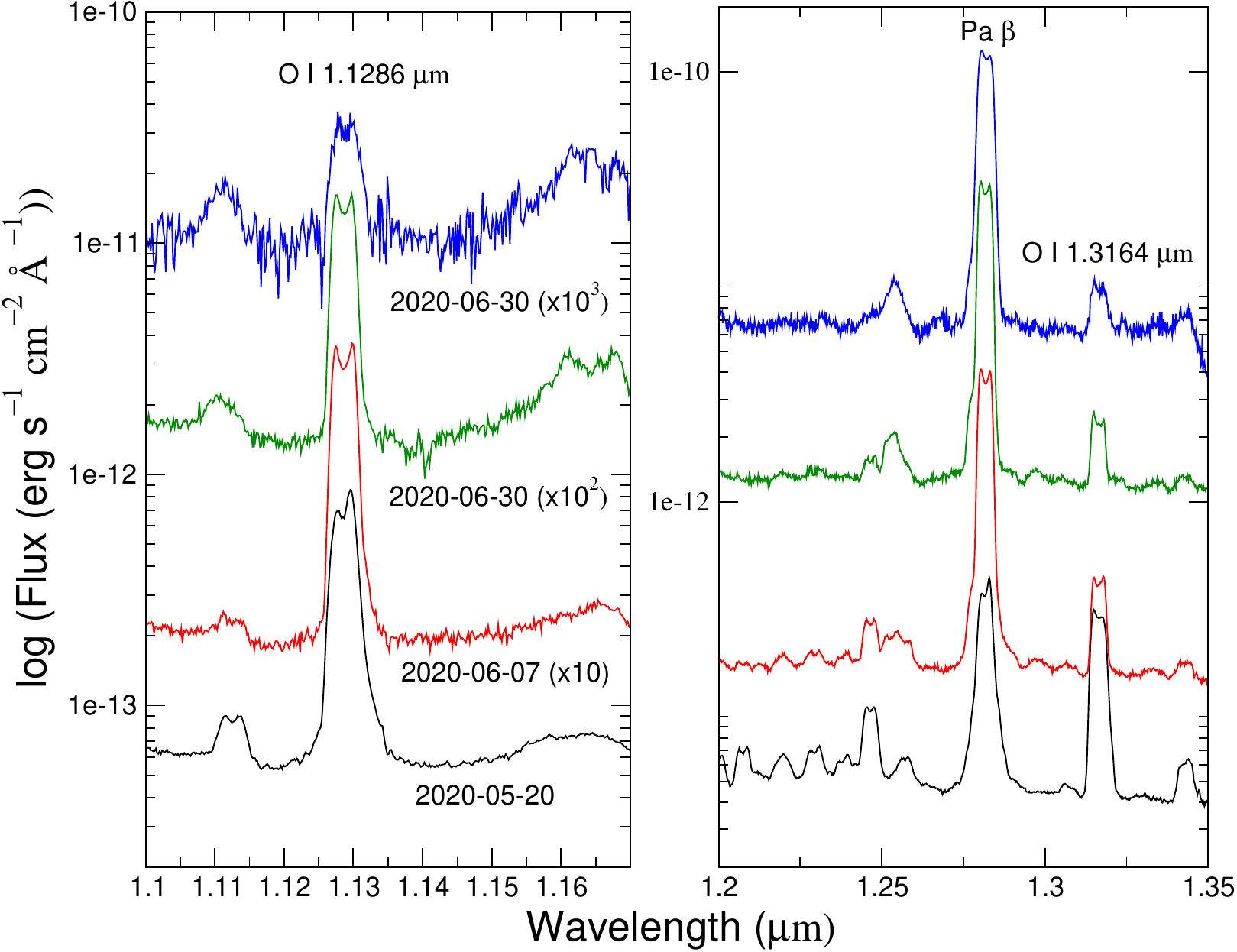}
    \caption{Evolution of the prominent Oxygen emission lines \ion{O}{i} 1.1286, and 1.3164 $\mu$m in the NIR spectrum of nova V2891 Cyg. The spectra are reddening corrected. See section~\ref{sec:oilines} for detailed discussion.}
    \label{fig:ir}
\end{figure}

The evolution of emission lines of \ion{O}{i} 7773, 8446 {\AA} in optical band, and \ion{O}{i} 1.1286, 1.3164 $\mu$m in NIR band are shown in Figure~\ref{fig:optical} and~\ref{fig:ir}, respectively. The behaviour of these emission features is explored in later sections in detail. The measured dereddened flux of various oxygen lines is given in Table~\ref{tab:line_flux}. The flux of \ion{H}{$\alpha$} is also given for comparison. In the early stages of an outburst, the spectrum was dominated by the \ion{H}{$\alpha$} and neutral emission features of oxygen at \ion{O}{i} 7773, and 8446 {\AA}. As nova ejecta expanded, the density decreased, and emission features from species with high ionisation potential started to appear in the spectrum. The emission lines of [\ion{O}{i}] at wavelengths 5577, 6300, and 6364 {\AA} were first seen in the spectra recorded between December 2019 and January 2020 (+84d to +122d) (see Figure 6 in \citet{2022MNRAS.510.4265K}). Following the method provided by \citet{2012AJWilliams}, \citet{2022MNRAS.510.4265K} utilised the these line fluxes to calculate the electron temperature of the neutral \ion{O}{i} regions to be around 5600K.
\par
The oxygen line of [\ion{O}{ii}] 7320 $\AA$ first appeared in the spectrum of 2020 May 06 (see \citet{2022MNRAS.510.4265K}). The flux of the line exhibited a consistent increase over time and was notably observed in the spectra observed over the period of October to November 2020. The emission line of [\ion{O}{iii}] 5007 {\AA} appeared on May 24, 2020 (+253d). It was certainly not present in the spectrum on May 20, 2020, \citep[see Figure 8 in][]{2022MNRAS.510.4265K}. Similar to the [\ion{O}{ii}] 7320 {\AA} line, this particular line exhibited an increase in flux with time and was prominently observed in spectra obtained from October to November 2020. The other forbidden lines, such as [\ion{N}{ii}] 5755 {\AA} line emerged in the optical spectra of 2020 May (+235d to +253d) along with the [\ion{O}{iii}] 5007 {\AA} which marked the onset of the nebular phase in V2891 Cyg \citep{2022MNRAS.510.4265K}.
\par
In the near-infrared domain, \ion{O}{i} 1.1286 $\mu$m and \ion{O}{i} 1.3164 $\mu$m were the two prominent \ion{O}{i} lines seen in novae spectra. While the production mechanism of \ion{O}{i} 1.1286 $\mu$m line is the same as the Lyman-$\beta$ fluorescence process for \ion{O}{i} 8446 {\AA} line (both the photons are produced in a single chain), the excitation mechanism for \ion{O}{i} 1.3164 $\mu$m emission line is suggested to be collisional excitation \citep{2012BASI...40..243B} in novae ejecta. The appearance and evolution of \ion{O}{i} 1.3164 $\mu$m emission is discussed in detail in Section~\ref{sec:OIlines_shock} in the context of V2891 Cyg.


\subsection{Photoionization perspective of the evolution of OI lines }\label{subsec-OxygenLines}

\begin{figure}

	\centering
	\includegraphics[scale=0.35]{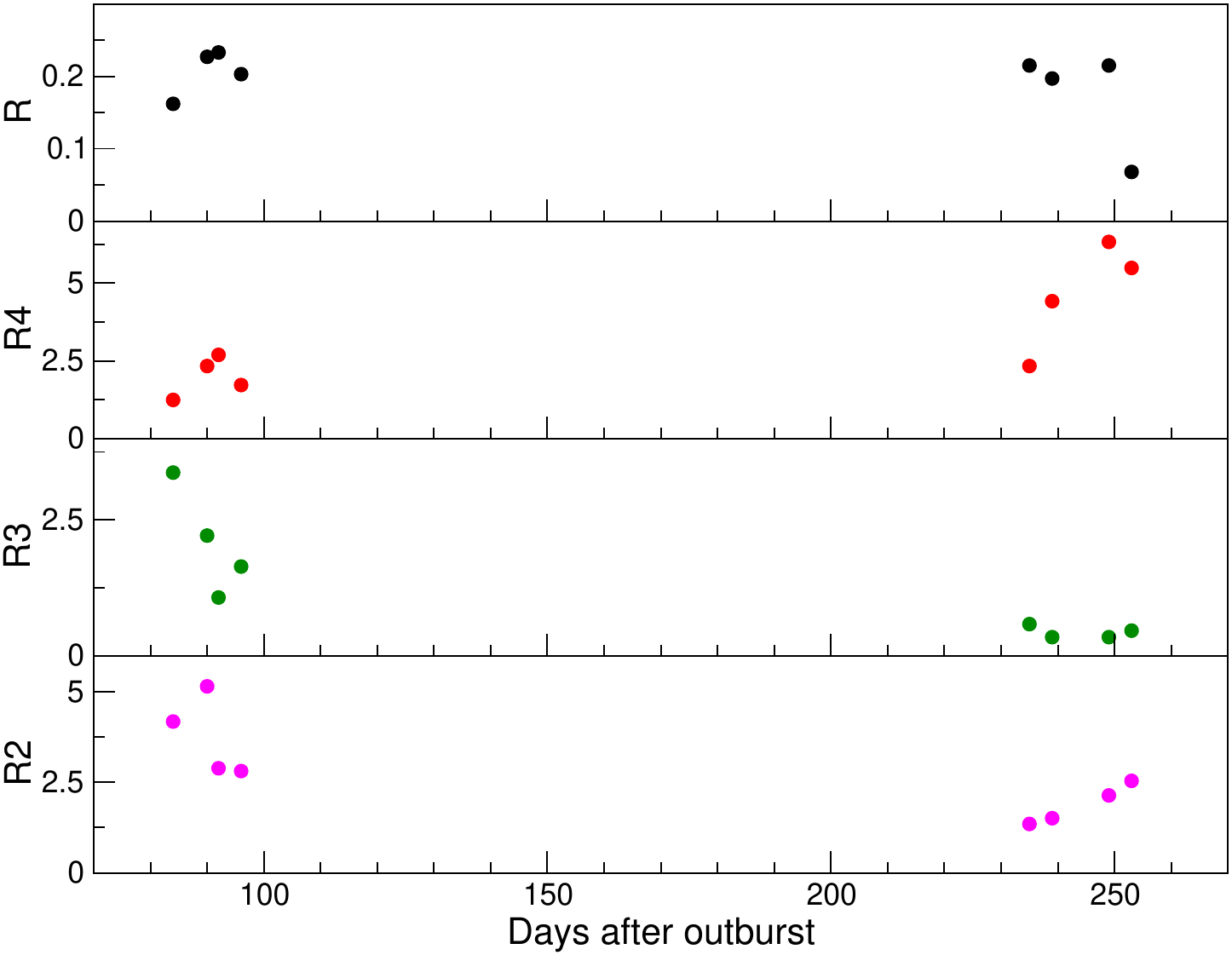}
	\caption{The ratio of Oxygen lines $R2=f(OI 8446)/f(OI 6300)$, $R3=f(OI 7773)/f(OI 6300)$, and $R4=f(OI 8446)/f(OI 7773)$. The ratio of $f(OI 8446)/f(H\alpha) = R$ is also shown in the top panel. See section~\ref{subsec-OxygenLines} for the discussion.}
	\label{fig-OxygenLineRatio}
\end{figure}

\citet{1995ApJSBhatia} and \citet{Kastner1995ApJ} provided a theoretical framework pertaining to the production of \ion{O}{i} lines in CNe. Various physical conditions were simulated to obtain the expected line ratios of prominent \ion{O}{i} lines under varying input parameters such as electron densities, temperature, photo-excitation rates, etc. The collisional excitation and Lyman $\beta$ fluorescence cases were also considered as excitation mechanisms for these emission lines. Following the nomenclature of \citet{Kastner1995ApJ}, we also evaluated the line ratios $R2 = f(\ion{O}{i} 8446)/f(\ion{O}{i} 6300)$, $R3 = f(\ion{O}{i} 7773)/f(\ion{O}{i} 6300)$ and $R4 = f(\ion{O}{i} 8446)/f(\ion{O}{i} 7773)$, for the reddening corrected fluxes, which are given in Table~\ref{table-OIRatio}. These are also plotted in 
Figure~\ref{fig-OxygenLineRatio}. Ratios ($R$) for eight epochs in 2019 Dec, and 2020 May are determined where all four \ion{O}{i} lines are recorded.

For higher electron densities and high photo-excitation rates, \citet{Kastner1995ApJ} showed that the log ($R2$), log ($R3$), and log ($R4$) would have positive values. These conditions were likely fulfilled in the 2019 Dec epoch ($\sim$74-105 days after the outburst), as shown by recombination analysis \citep{2022MNRAS.510.4265K} for electron densities. The ratio $R4$ should have been 0.6 under normal recombination in optically thin conditions \citep{2008A&AMunari}. However, it was found to be in the range of 1.24–2.69, which is attributed to the higher strength of the \ion{O}{i} 8446 $\AA$ line due to the Lyman-$\beta$ fluorescence. This requires a large optical depth in \ion{H}{$\alpha$}. During 2019 Nov-Dec, the ratio of $f(\ion{O}{i} 8846)/f(\ion{H}{$\alpha$})$ was $\sim$ 0.1-0.2 while it was expected to be as low as $10^{-3}$ in optically thin conditions \citep{Strittmatter1977,2008A&AMunari}. The higher values of this ratio clearly indicate a large optical depth in \ion{H}{$\alpha$}, again indicating the ejecta's dense nature. The $f(\ion{O}{i} 8846)/f(\ion{H}{$\alpha$})$ ratio remained around 0.2 during early May 2020. However, the spectrum of 2020 May 24 showed a sharp drop to 0.07. In the later evolution (2020 Oct–Nov), the ratio further dropped to 0.007 (see Table~\ref{tab:line_flux}). The ratio $R4$, on the other hand, went up from 2.33 to 6.33 between 2020 May 6 and 20. It then dropped to 5.52 on 2020 May 24. The \ion{O}{i} 8446 {$\AA$} and 7773 {$\AA$} emissions were below the detection limits in the spectra recorded in late 2020 Nov-Dec 2020.
\par 
It is imperative to recall that, in the optical domain of novae spectra, the \ion{O}{i} emission at $8446$ {$\AA$} ($3s^3S^0-3p^3P$) is a prominent feature. Apart from novae, many other astronomical objects, such as \ion{H}{ii} regions, planetary nebulae, Orion nebula, Seyfert galaxies, etc., also show the emission from \ion{O}{i} $\lambda 8446$ {$\AA$} \citep[for e.g.,][and references therein]{Grandi1975,Grandi1976ApJ,Strittmatter1977,Grandi1980,Rudy1989}. This high observed strength of this line is not only due to the usual physical processes of recombination or electron collisions but also to the fluorescence, which has also been proposed to be a dominant mechanism. \citet{Grandi1975} discussed the two possible fluorescence mechanisms which could enhance the strength of \ion{O}{i} $8446$ {$\AA$} emission line; fluorescence induced by the continuum of the star exciting the nebula and resonance fluorescence of \ion{O}{i} by Lyman $\beta$. In the case of novae, it is proposed that the primary excitation mechanism for the \ion{O}{i} $ 8446$ {$\AA$} line is the Bowen mechanism of Ly$\beta$ fluorescence, which arises due to the photoexcitation of the \ion{O}{i} resonance line at $ 1025.77$ {$\AA$} ($2p^4(^3P_2)-3d^3D^0$) by the hydrogen Ly$\beta$ line at $ 1025.72$ {$\AA$}. The pumping of \ion{O}{i} $ 1025.77$ {$\AA$} ground state due to the close proximity between these two lines populates the $3d^3D^0$ level of \ion{O}{i}, leading to a downward cascade of transitions that produces the emission lines at $1.1286~\mu$m ($3p^3P-3d^3D^0$), $ 8446$ {$\AA$} ($3s^3S^0-3p^3P$), and $ 1304$ {$\AA$} ($2p^3P-3s^3S^0$) (see Figure 1 in \citet{Kastner1995ApJ}). Thus, it is not unlikely that the high flux of \ion{O}{i} 8446 {$\AA$} emission line in V2891 Cyg was due to the same fluorescence mechanism.
\par
While the \ion{O}{i} 8446 {$\AA$} line showed behaviour consistent with the fluorescence excitation mechanism, it was intriguing to notice the flux and evolution of the \ion{O}{i} 7773 {$\AA$} emission line, which will be discussed in later sections. The flux of the \ion{O}{i} 7773 {$\AA$} line was comparable to that of the \ion{O}{i} 8446 {$\AA$} during the early spectral evolution (November-December 2019) of the V2891 Cyg. A P-Cygni profile was also observed in both emission lines in the spectrum dated December 19, 2019. The \ion{O}{i} 7773 {$\AA$} transition is characterized as a quintet, while the 8446 {$\AA$} transition is a triplet transition \citep{2012AJWilliams}. The behaviour of \ion{O}{i} 7773 $\AA$ and 8446 $\AA$ lines and the correlation in their fluxes are different in the He/N and \ion{Fe}{ii} class of novae \citep{2012AJWilliams}. The emission line of \ion{O}{i} 8446 {$\AA$} is usually present in the early decline spectra of all novae, while the emission line of \ion{O}{i} 7773 {$\AA$} is occasionally detected in the early stage spectrum of \ion{Fe}{ii} class novae, as the transition may not be formed in the low-density gas \citep{2012AJWilliams}. In the early evolution of V2891 Cyg, \ion{O}{i} 7773 $\AA$ line showed comparable strength to that of the \ion{O}{i} 8446 $\AA$ line, suggesting a high-density region in the ejecta from where \ion{O}{i} 7773 $\AA$ emission was originated. The flux of the \ion{O}{i} 7773 {$\AA$} transition is primarily determined by collisional excitation and recombination, and it is not expected to have a flux comparable to that of the fluorescence-excited \ion{O}{i} 8446 {$\AA$} emission line. However, the strength of \ion{O}{i} 7773 {$\AA$} can be increased through collisional de-excitation originating from the $3p^{3}P$ upper level of the \ion{O}{i} 8446 {$\AA$} transition to the $3p^5P$ upper level of \ion{O}{i} 7773 {$\AA$} \citep{Kastner1995ApJ,2012AJWilliams}. Theoretical models by \citet{Kastner1995ApJ} suggest that the explanation of the significant emission equivalent widths of \ion{O}{i} 8446 {$\AA$} in novae requires a significant Ly $\beta$ flux and relatively low kinetic temperatures, often around T $\sim$ 7000 K, as these characteristics are frequently detected in the spectra during the initial stages of decline. These variables play a pivotal role in attaining elevated rates of photoexcitation to the \ion{O}{i} $3p^3P$ energy level \citep{Kastner1995ApJ}. Later, as the nebular phase began to commence in May 2020, the \ion{O}{i} 8446 {$\AA$} emission line started to gain strength with respect to \ion{O}{i} 7773 {$\AA$} emission (Table~\ref{table-OIRatio}).
\par 
The forbidden lines of neutral oxygen [\ion{O}{i}] at 5577, 6300, 6364 $\AA$ usually appear earlier in the spectral evolution due to their high critical densities. In some cases, they persist even in the presence of dominant [\ion{O}{iii}] emission and other high ionisation lines. \citet{1994ApJSWilliams} suggested that neutral oxygen comes from small, dense condensations of neutral material that are embedded in ionised ejecta in the ambient environment. This would explain the large optical depth in the ratio of the 6300 and 6364 $\AA$ lines. Their flux ratio can also be used to estimate the photo-excitation rate from the data given in \citet{Kastner1995ApJ}. The recombination analysis of the nova ejecta on 2020 June 7 (267 days after the outburst) showed that the ejecta were in an optically thin condition with electron density $N_e \sim 10^8$ cm$^{-3}$ and electron temperature $T_e \sim 10000$ K \citep{2022MNRAS.510.4265K}. Assuming the temporal dependence of electron density follows the profile $N_e \propto d^{-3} \propto t^{-3}$, where d is the distance travelled by the ejecta in $t$ time, this translates into $N_e \sim 10^{9} \text{cm}^{-3}$ in December 2019 (75-105 days after the outburst). At these densities, the observed ratio log ($R2$) of $\sim$0.55 in Dec 2019 can be interpolated for photo-excitation rates of $10^{-3} - 10^{-2} \text{s}^{-1}$ for $T_e$ between 5000 and 10000 K. However, at these temperatures and photo-excitation rates, the ratio log ($R3$) was rather high. It was expected to be in the range of $-0.58$ to $-2.05$, for $T_e$ between 5000-10000 K and a photo-excitation rate of $10^{-2} \text{s}^{-1}$. Similarly, the observed ratio log ($R4$) was relatively lower. The expected range, for $N_e \sim 10^{9} \text{cm}^{-3}$, is between 2.64-4.74 for the photo-excitation rates of $10^{-2} \text{s}^{-1}$ and 1.00-4.72 for the photo-excitation rates of $10^{-4} \text{s}^{-1}$. Thus, the flux of \ion{O}{i} 7773 $\AA$ emission was exceedingly high to determine the expected photo-ionization rate in 2019 Dec. The log ($R2$) was in the range of 0.45-0.62 for Dec 2019 and 0.13-0.41 in May 2020. For the density $N_e \sim 10^{9} \text{cm}^{-3}$ in Dec 2019 and $\sim 10^{8} \text{cm}^{-3}$ in May 2020, the ratios correspond to interpolated photo-excitation rates of $\sim 10^{-3} - 10^{-2} \text{s}^{-1}$ and $\sim 10^{-3.3} - 10^{-2.3} \text{s}^{-1}$, respectively, for $T_e$ in the range of 5000-10000K.

\subsection{The observed characteristics of O \textsc{i} 7773 {\AA} and O \textsc{i} 1.3164 $\mu$m emission lines}\label{subsec:7773and13164}

Out of several oxygen lines seen in the evolution of the optical and NIR spectra of V2891 Cyg, \ion{O}{i} 7773 {\AA} and \ion{O}{i} 1.3164 $\mu$m lines exhibited some very interesting evolutionary characteristics. The appearance, disappearance, and re-appearance of P-Cygni profiles of \ion{O}{i} 7773 {\AA} in the spectra of V2891 Cyg was a critical feature that was discussed in the previous work \citep{2022MNRAS.510.4265K}. The synergy between P-Cygni appearance and brightness peaks in the nova light curve led the authors to propose that the nova outburst consists of multiple episodes of periodic mass ejection. The typical speed of these mass ejections, which is usually correlated with the absorption dip in the blue part of the P-Cygni profile, was found to be 300-700 km s$^{-1}$. For a detailed discussion of concerned observations, please see \citet{2022MNRAS.510.4265K} (Section~\ref{sec:o1_7773}).


\begin{landscape}
\centering
\begin{table}
\caption{Reddening corrected flux ($\times10^{-11}$erg s$^{-1}$ cm$^{-2}$) of oxygen lines and \ion{H}{$\alpha$} during the evolution of nova V2891 Cyg}
\centering
\setlength{\tabcolsep}{8pt}
\renewcommand{\arraystretch}{1.2}
	\begin{tabular}{lc cc cc cc cc cc cc}
           \hline
		\hline
		Date of     & Days &  [\ion{O}{iii}] 4959 {\AA}  & [\ion{O}{iii}] 5007 {\AA} & [\ion{O}{i}] 5577 {\AA} & [\ion{O}{i}] 6300 {\AA} & [\ion{O}{i}] 6364 {\AA} &	\ion{H}{$\alpha$}  & \ion{O}{i} 7773 {\AA} & \ion{O}{i} 8446 {\AA} & \ion{O}{i} 1.1286 $\mu$m & \ion{O}{i} 1.3164 $\mu$m  \\
		Observation &         &    &          &            &           &            &        		&          &          &           &             \\
		            &           &          &            &           &            &        		&          &          &           &             \\
		\hline
		2019-11-01 & 48 &  ...               &  ...               & ...            & ...             &  ...  
		&   14.45$\pm$0.13  &  1.18$\pm$0.02  &  1.50$\pm$0.02  &  ...  &  ...  & \\
		2019-11-11 &  58 & ...                &  ...              & ...             & ...             &  ... 
		&   14.75$\pm$0.13  &  1.63$\pm$0.03  &  ...    &  ...  &  ...  &          \\
		2019-11-12 & 59 &  ...                &  ...               & ...             & ...             &  ...  
		&   10.96$\pm$0.11  &  1.39$\pm$0.03  &  1.10$\pm$0.02 &  ...  &  ...  & \\
		2019-11-19 & 66 &  ...                &  ...               & ...             & ...             &  ...  
		&   26.72$\pm$0.20  &  1.41$\pm$0.05  &  4.16$\pm$0.05 &  ...  &  ...  & \\
		2019-12-01 & 78 &  ...                &  ...               & ...             & ...            &  ...  
		&   16.08$\pm$0.17  &  ...              &  ...         &  ...  &  ...  &     \\
		2019-12-07 & 84 & ...                &  ...               & 4.46$\pm$0.45 & 5.01$\pm$0.15 & 
		2.60$\pm$0.16 &  128.92$\pm$0.32  & 16.91$\pm$0.14  & 20.96$\pm$0.11  &  ...  &  ...  &\\
		2019-12-13 &  90 & ...                &  ...               & ...             & 6.00$\pm$0.35 & 
		2.11$\pm$0.30 &  135.58$\pm$0.86  & 13.27$\pm$0.10  & 30.91$\pm$0.10  &  ...  &  ...  &\\
		2019-12-15 &  92 & ...                &  ...               & 3.60$\pm$0.41 & 4.99$\pm$0.15 & 
		1.99$\pm$0.14 &   61.78$\pm$0.26  &  5.38$\pm$0.10  & 14.45$\pm$0.10 &  ...  &  ...  & \\
		2019-12-19 & 96 &  ...                &  ...               & ...             & 2.35$\pm$0.18 & 
		1.65$\pm$0.19 &   32.55$\pm$0.26  &  3.85$\pm$0.07  &  6.61$\pm$0.06  &  ...  &  ...  &\\
		2020-01-04 & 112 &  ...                &  ...               & ...             & 2.34$\pm$0.21 &  ...  
		&   23.62$\pm$0.22  &  3.53$\pm$0.04  &  ...         &  ...  &  ...  &     \\
		2020-01-13 & 121 & ...                &  ...               & ...             & ...             &  ...  
		&   30.94$\pm$0.67  &  4.07$\pm$0.06  &  ...          &  ...  &  ...  &    \\
		2020-05-06 &  235 & ...                &  ...               & ...             & 4.92$\pm$0.21 & 
		2.26$\pm$0.20 &   30.77$\pm$0.27  &  2.85$\pm$0.12  &  6.63$\pm$0.10  &  ...  &  ...  &\\
		2020-05-10 & 239 &  ...                &  ...               & ...             & 4.84$\pm$0.18 & 
		2.92$\pm$0.19 &   37.01$\pm$0.29  &  1.66$\pm$0.04  &  7.32$\pm$0.05  &  ...  &  ...  &\\
		2020-05-20 & 249 &  ...                &  ...               & ...             & 3.83$\pm$0.19 & 
		2.19$\pm$0.20 &   38.08$\pm$0.30  &  1.29$\pm$0.05  &  8.19$\pm$0.05 & 3.08$\pm$0.06 & 1.43$\pm$0.04  \\
		2020-05-24 &  253 & ...                &  29.44$\pm$1.43  & ...             & 2.28$\pm$0.09 & 
		1.23$\pm$0.08 &   84.39$\pm$0.23  &  1.05$\pm$0.02  &  5.77$\pm$0.02  &  ...  &  ...  &\\
            2020-06-07 & 267 & ...               &  ...               & ...            & ...          &  ...  
		&   ...  &  ...  &  ...  & 1.28$\pm$0.10& 0.12$\pm$0.04 \\
            2020-06-30 &  290 &...               &  ...               & ...            & ...          &  ...  
		&   ...  &  ...  &  ... & 0.54$\pm$0.03 & 0.05$\pm$0.01\\
            2020-09-19 & 371 &...               &  ...               & ...            & ...          &  ...  
		&   ...  &  ...  &  ...  & 0.09$\pm$0.01 & 0.02$\pm$0.008\\
		2020-10-23 & 375 & ...                &  ...               & ...             & ...             &  ...  
		&   16.37$\pm$0.12  & ...               &  0.04$\pm$0.01    &  ...  &  ...  &  \\
		2020-11-01 &  414 & ...                &  ...               & ...             & ...            &  ...  
		&   14.08$\pm$0.22  & ...               &  0.25$\pm$0.03    &  ...  &  ...  &  \\
		2020-11-02 & 415 &  3.09$\pm$0.49 &  8.57$\pm$0.24  & ...            & ...            & 
		...             &    7.56$\pm$0.03  &  ...  &  0.05$\pm$0.01    &  ...  &  ...  &  \\
		2020-11-03 &  416 & ...           &   7.01$\pm$0.49  & ...             & ...            &  ...  
		&    6.33$\pm$0.06  & ...              & ...             &  ...  &  ...  &      \\
		2020-11-16 &  429 & 5.82$\pm$0.61 &  12.29$\pm$0.11  & ...             & ...             & 
		...             &    8.32$\pm$0.04  & ...               & ...            &  ...  &  ...  &       \\
		2020-11-17 &  430 & 20.37$\pm$6.26 &  75.71$\pm$5.33  & ...             & ...            & 
		...            &   31.03$\pm$0.27  & ...               & ...             &  ...  &  ...  &     \\
		2020-11-19 & 432 &  2.77$\pm$0.84 &   8.75$\pm$0.47  & ...             & ...             & 
		...             &    6.87$\pm$0.05  & ...              & ...            &  ...  &  ...  &       \\
		2020-11-27 & 440 &  ...               &  ...               & ...            & ...            &  ... 
		&   15.28$\pm$0.27  & ...              & ...          &  ...  &  ...  &        \\
		2020-11-29 & 442 & 10.14$\pm$1.53 &  35.38$\pm$1.06  & ...             & ...            & 
		...             &   15.13$\pm$0.12  & ...              & ...         &  ...  &  ...  &          \\
		2020-12-08 &  451 & ...               &   7.15$\pm$0.75  & ...            & ...            &  ...        
		&    4.43$\pm$0.08  & ...              & ...          &  ...  &  ...  &        \\
		\hline
		\hline
	\end{tabular}
        \label{tab:line_flux}
\end{table}
\end{landscape}


\begin{table*}
	\centering
 \setlength{\tabcolsep}{12pt}
	\caption{Ratio of the reddening corrected fluxes of \ion{O}{i} lines. $R2=f(OI 8446)/f(OI 6300)$, $R3=f(OI 7773)/f(OI 6300)$, and $R4=f(OI 8446)/f(OI 7773)$.}
	\begin{tabular}{lccc}
		\hline
		Date of Observation  & $R2$ & $R3$ &$R4$ \\
		(Days after Outburst)& (log $R2$) & (log $R3$) & (log $R4$)  \\
		\hline
		\hline		 								
		2019-12-07 (84)  & 4.18(0.62) $\pm$ 0.12 & 3.37(0.53) $\pm$ 0.10   &1.24 (0.09) $\pm$ 0.01 \\ 
		2019-12-13 (90)  & 5.15(0.71) $\pm$ 0.30 & 2.21(0.34) $\pm$ 0.13   &2.33 (0.37) $\pm$ 0.01 \\ 
		2019-12-15 (92)  & 2.89(0.46) $\pm$ 0.08 & 1.07(0.03) $\pm$ 0.03   &2.69(0.43) $\pm$ 0.05  \\ 
		2019-12-19 (96)  & 2.81(0.45) $\pm$ 0.21 & 1.64(0.22) $\pm$ 0.12   &1.72 (0.24) $\pm$ 0.03 \\ 
		2020-05-06 (235) & 1.35(0.13) $\pm$ 0.06 & 0.58(-0.24) $\pm$ 0.03  &2.33(0.37) $\pm$ 0.10 \\ 
		2020-05-10 (239) & 1.51(0.18) $\pm$ 0.05 & 0.34(-0.47) $\pm$ 0.01  &4.42(0.65) $\pm$ 0.11 \\ 
		2020-05-20 (249) & 2.14(0.33) $\pm$ 0.10 & 0.34(-0.47) $\pm$ 0.02  &6.33(0.80) $\pm$ 0.24 \\ 
		2020-05-24 (253) & 2.54(0.41) $\pm$ 0.10 & 0.46(-0.34) $\pm$ 0.02  &5.49(0.74) $\pm$ 0.10 \\ 		
		\hline
		\hline
	\end{tabular}
	\label{table-OIRatio}
\end{table*}


The time-varying P-Cygni profiles in nova ejecta have been discussed in the literature as well. For example, \citet{2011PASJTanaka} investigated six CNe, namely V1186 Sco, V2540 Oph, V4745 Sgr, V5113 Sgr, V458 Vul, and V378 Ser. The authors discovered identical occurrences of re-brightening and re-appearance of P-Cygni profiles in various emission lines. During the early stages of these novae, when they reached their maximum brightness, the emission lines had a P-Cygni profile. The absorption component of the P-Cygni profiles subsequently disappeared following the maximum. Nevertheless, upon the novae's subsequent increase in brightness to reach the next maximum, the P-Cygni profiles resurfaced, indicating the presence of the absorption component. The repeated appearances of P-Cygni profiles observed in the emission lines at the rebrightening provide evidence of many instances of mass ejection in classical novae \citep{2020ApJAydi}. 
\par 
It is noteworthy that the time variation of the P-Cygni profile was only observed in the \ion{O}{i} 7773 {\AA} line in the case of V2891 Cyg. While in the earliest epochs, the P-Cygni profile was noted in other lines, such as \ion{O}{i} 8446 {\AA}, \ion{H}{$\alpha$}, etc., which later disappeared as expected due to the usual spectral evolution caused by the expansion of the ejecta. The profile of the \ion{O}{i} 7773 {\AA} line stands out, indicating a plausible stratification of oxygen in this particular case, leading to different origins of these lines within the nova ejecta. As a contrary example, it is interesting to note the spectral evolution of oxygen lines in the ejecta of nova SN 2010U \citep{2013ApJCzekala} where both the \ion{O}{i} 7773 {\AA} and 8446 {\AA} lines showed the presence of the P-Cygni profile and exhibited coherent evolution. The observed differences in the line profiles and their evolution in several \ion{O}{i} lines in nova V2891 Cyg might suggest that these lines originated in different regions in the nova ejecta with varying physical and chemical conditions. These aspects are explored in detail in Section~\ref{sec:cloudy}.
\par
\begin{figure}
    \centering
    \includegraphics[scale=0.6]{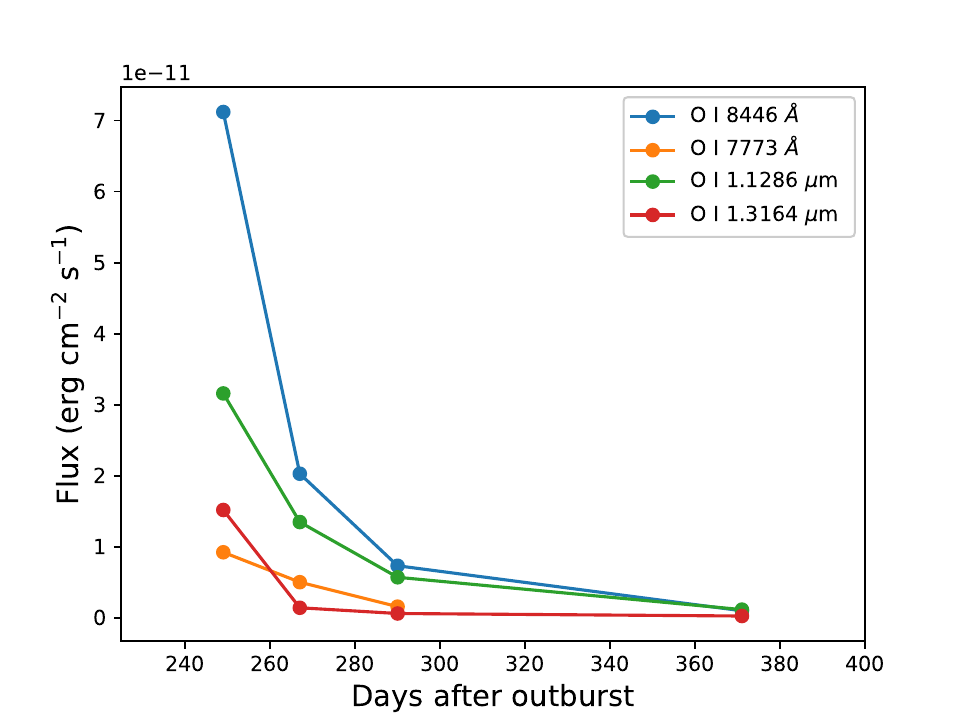}
    \caption{The evolution of absolute reddening-corrected observed flux for the prominent neutral oxygen lines \ion{O}{i} 7773 {\AA}, 8446 {\AA}, 1.1286 $\mu$m, and 1.3164 $\mu$m.}
    \label{fig:OiNirflux}
\end{figure}

In the NIR domain of V2891 Cyg, the evolution of the line flux of the \ion{O}{i} 1.3164 $\mu$m line was a notable feature. The variation observed in this line is shown in Figure~\ref{fig:ir}. The emission of \ion{O}{i} 1.1286 $\mu$m is also presented, showing a contrasting behaviour characterized by the maintenance of its flux compared to \ion{O}{i} 1.3164 $\mu$m emission line. Table~\ref{tab:line_flux} presents the observed dereddened flux of the \ion{O}{i} 1.3164 $\mu$m line and the \ion{O}{i} 1.1286 $\mu$m line in the NIR spectrum. The evolution of their dereddened observed flux with the number of days after the outburst is shown in Figure~\ref{fig:OiNirflux}. A significant decline in the measured flux was found in the \ion{O}{i} 1.3164 $\mu$m line when compared to other \ion{O}{i} lines. It is of extreme interest to note that such a sharp drop in the line flux of \ion{O}{i} 1.3164 $\mu$m was very well aligned with the short time span of dust formation and when coronal lines began to appear in the observed spectrum \citep{2022MNRAS.510.4265K}. It was argued by \citet{2022MNRAS.510.4265K} that the dust formation in this nova was most likely due to shocks, which could have been entirely possible due to the conditions of multiple mass ejections with suitable velocity differences during the nova outburst. Now, the apparent coincidence of the flux variation of the \ion{O}{i} 1.3164 $\mu$m line during the dust formation epoch provides us with another opportunity to revisit the conditions of dust formation using these oxygen emission lines as the probe. In the following sections, we will examine the possible factors contributing to this behaviour to better understand the physical mechanisms accountable for such behaviour in the \ion{O}{i} 1.3164 $\mu$m line during the evolution of V2891 Cyg.

\section{Nebular phase analysis}\label{sec:cloudy}
 
Numerous shock-induced processes, such as gamma-ray emission, X-ray, etc., are known and typically observed in the early evolution of novae outbursts \citep[see][and references therein]{Chomiuk2014Natur,Vlasov2016MNRAS,Metzger2016MNRAS,Li2017NatAs,Chomiuk2021ARA&A}. Additionally, dust formation in novae outbursts has also been a well-established phenomenon. However, the role of shock in dust formation is one of the most intriguing concepts discussed in the literature. Recently, \cite{Metzger2016MNRAS} and \citet{Derdzinski2017MNRAS} provided a theoretical framework for shock-induced dust formation in novae. While there are studies on shock-induced formation dust formation in supernovae \citep{2018ApJSarangi}, such studies and observations are relatively rare for novae. The behaviour of \ion{O}{i} lines during the evolution of nova V2891 Cyg is a suitable opportunity to explore the physics of the aforementioned phenomena in depth by employing a phenomenological approach. 
\par
Nova V2891 Cyg entered the optically thin nebular phase during May 2020 \citep{2022MNRAS.510.4265K}. The optically thin spectrum forms under significantly less complicated conditions than the optically thick phase. At this point, the spectra could be well modelled with a photoionization code that encompasses the major heating and cooling mechanisms as well as a significant number of predicted emission lines \citep{1998PASP..110....3G,2001MNRASSchwarz}. In the present study, we used photoionization modelling to investigate the nebular phase spectra of V2891 Cyg. We derived the physical and chemical conditions of the ejecta and their evolution as the system unfolds. A total of four epochs of optical/NIR spectra of this nebular phase are used, namely, 2020 May 20 (+249 d), 2020 June 07 (+267 d), 2020 June 30 (+290 d), and 2020 September 19 (+371 d). These epochs will be referred to as epoch 1, epoch 2, epoch 3, and epoch 4 hereafter.


\subsection{Photoionization modelling}\label{subsec:photo}

\textsc{cloudy} is a widely utilised photoionization code that is employed to simulate the physical and chemical properties and consequent emission spectra of ionized gas clouds. \textsc{cloudy} simulates the physical circumstances of non-equilibrium gas clouds subjected to an external radiation source using detailed microphysics. For a specified set of input parameters, it concurrently solves the equations of the thermal, statistical, and chemical equilibrium for a model emission nebula incorporating major ionization and recombination processes in a self-consistent algorithm and generates the output spectrum. For a detailed discussion on the \textsc{cloudy}, please see \citet[][and references therein]{2017RMxAA..53..385F}. The \textsc{cloudy} has been extensively utilised in numerous investigations to study novae \citep[see, e.g.][and references therein]{2001MNRASSchwarz,2002ApJSchwarz,2003AJShore,2010AJHelton,2019A&APavana,2019MNRAS.483.4884M,2022ApJPandey,2022MNRASPandey,2024MNRASHabtie}
\par 
\citet{2022MNRAS.510.4265K} estimated the temperature of the pseudo-photosphere during the nebular phase in the range of 35000–54000 K. Coronal emission was also seen in these epochs, and they argued against the presence of a hot central photoionization source based on Swift Target of Opportunity (ToO) observations, utilising the X-ray Telescope (XRT) and UV instrument onboard (target ID 12868), which failed to detect any UV/X-ray emission.\citet{2022MNRAS.510.4265K} then presented a case in favour of shock-induced coronal emission. They also estimated electron densities during these nebulae epochs using Case-B recombination analysis of hydrogen Br lines. These estimates were taken as the initial parameters to construct the preliminary phenomenological model of V2891 Cyg.
\par 
We began by assuming a central ionising source whose parameters are determined by a blackbody of the aforementioned temperature and input luminosity (erg/s). A spherically symmetric ejecta with dimensions determined by the inner and outer radii surrounds this central ionizing source. The central ionising source provides heat to the surrounding ejecta by photoionization. The size of the ejecta was estimated using the minimum and maximum expansion velocities of the ejecta, determined by the full width at half-maximum (FWHM) of emission lines \citep{2022MNRAS.510.4265K} and the time elapsed between the nova outburst and discovery for each epoch.
\par
The ejecta density in \textsc{cloudy} models is set by the total hydrogen number density, $n(\text{H})$ [cm$^{-3}$] parameter, given by,
\begin{equation}
n(\text{H}) = n(\text{H}^{0}) + n(\text{H}^{+}) + 2n(\text{H}_{2}) + \sum_{\text{other}} n(\text{H}_{\text{other}})~ \text{cm}^{-3},
\end{equation}
where, $n(\text{H}^{0})$, $n(\text{H}^{+})$, $ 2n(\text{H}_{2})$, and $n(\text{H}_{\text{other}})$ represent hydrogen in neutral, ionized, molecular, and all other hydrogen bearing molecules, respectively. We used a radius dependent power-law hydrogen density distribution, $\rho \propto r^{\alpha}$, given by,
\begin{equation}\label{eq:nh}
  n(r)=n(r_{\text{in}})(r/r_{\text{in}})^{\alpha},
\end{equation}
where $n(r)$ and $n(r_{\text{in}})$ represent the ejecta density at distance $r$ and the inner radius $(r_{\text{in}})$, respectively. According to spherically symmetric, line blanketing non-LTE expanding atmospheric models of Novae \citep[see][references therein]{hauschildt_2008,2001ApJ...547.1057S,1996ApJ...462..386H}, during the early phases of nova ejecta evolution, the density profile is highly sharp (with $\alpha \sim -10$). The temperature of the ejecta decreases gradually, from approximately 15,000 K to below 10,000 K, due to adiabatic expansion. The process of expansion also leads to a reduction in optical depth, thereby enabling the deeper layers of the ejecta to become perceptible. As a result, the density distribution of the ejected material exhibits a decrease in steepness (with $\alpha \sim -3$). This density profile of the nova atmosphere suggests a significant geometric expansion, resulting in a diverse range of temperatures and ionisation states in both the continuum and line-forming regions \citep{hauschildt_2008}. Thus, we chose $\alpha = -3$ for all \textsc{cloudy} models.
\par
The filling factor describes the volume percent occupied by gas. All \textsc{cloudy} models use a radial dependant filling factor, which is given by,

\begin{equation}\label{eq:filling}
f(r)=f(r_{\text{in}})(r/r_{\text{in}})^{\beta}.
\end{equation}

The filling factor in novae ejecta has small values, which lie in the range from 0.01 to 0.1 \citep{shore_2008}. We set the values of f = 0.1 and $\beta = 0$ for V2891 Cyg. The \textit{abundance} parameter in the \textsc{cloudy} determines the abundance of ejecta. All \textsc{cloudy} models presented here vary elemental abundances of those species whose emission lines are observed in the spectra. The values of the other elements were kept at their solar values from \citet{2010Ap&SS.328..179G}. Similar values of input parameters are also used in a variety of studies, such as V1974 Cyg \citep{Valandingham2005ApJ}, V4160 Sgr \citep{2007ApJ...657..453S}, V1065 Cen \citep{2010AJHelton}, RS Oph 2006 \citep{2018MNRASMondal}, T Pyx \citep{Pavana2019A&A}, RS Oph 2021 \citep{2022MNRASPandey}, V1280 Sco \citep{2022ApJPandey}, V1674 Her \citep{2024MNRASHabtie} etc.
\par
Temperature and luminosity of the central ionising source, inner and outer radii of the spherically symmetric ejecta, and chemical abundances of the ejecta were the input parameters for our models. To compute the set of synthetic spectra, we varied the values of all input parameters simultaneously in smaller increments throughout a large sample space. These synthetic spectra were then compared to the observed spectra to determine the best-fit model parameters (Table~\ref{tab:parameters}).
\par
A single-uniform density model could not produce the observational spectral features of high ionisation potential. An additional low-density component was added to the model. Several other studies have also demonstrated the limits of the one-component model in a variety of cases, such as V1974 Cyg \citep{Valandingham2005ApJ}, V4160 Sgr \citep{2007ApJ...657..453S}, RS Oph 2006 \citep{2018MNRASMondal}, V1065 Cen \citep{2010AJHelton}, RS Oph 2021 \citep{2022MNRASPandey}, and V1280 Sco \citep{2022ApJPandey}, T Pyx \citep{Pavana2019A&A}, etc. The physical parameters of both components were identical, except for the hydrogen density of the ejecta and their covering factors, which we varied independently. The covering factors of both components were adjusted so that their sum was less than or equal to 1. The final spectrum was produced by multiplying the spectrum of each component by their respective covering values and then adding them. We used spectrum obtained from the continuum file generated by the \textsc{cloudy}.
\par
While this two-component model had been successful in generating the majority of emission lines and continuum, it could not generate the most interesting \ion{O}{i} 7773 {\AA} and \ion{O}{i} 1.3164 $\mu$m emission lines, as we described in section~\ref{subsec:7773and13164}. Therefore, a third component of very high density and low mass was added to the model, signifying the dense clumps with high O abundance for \ion{O}{i} 7773 {\AA} emission. This is discussed in detail in section~\ref{sec:o1_7773}. The schematic of the composite phenomenology is shown in Figure~\ref{fig:model}.
\par 
To find the best \textsc{cloudy} model, we used statistical methods like chi-square ($\chi^2$) and reduced chi-square ($\chi^2_{red}$), which are given by the following:

\begin{equation}
\chi^2 =\sum_{i=1}^n\frac{(M_i-O_i)^2}{\sigma_i^2},
\end{equation}

where $O_i$, $M_i$, $n$, and $\sigma_i$ represent observed line flux ratios, modelled line flux ratios, number of emission lines used in the model, and error in the observed line flux ratios, respectively. The reduced $\chi^{2}$ is given by

\begin{equation}\label{eq:redchisq}
\chi^{2}_{\text{red}} = \frac{\chi^2}{\nu},
\end{equation}

where $\nu$ represents the number of degrees of freedom (DOF), which is determined by the difference between the number of observed emission lines ($n$) and the number of free parameters ($n_p$), i.e., $\nu$ = $n - n_p$. For an appropriate fit, the values of $\chi^2 \sim \nu$ and $\chi^2_{red}$ should be low, usually between 1 and 2.
\par
\begin{figure}
\centering
  \includegraphics[scale=0.35]{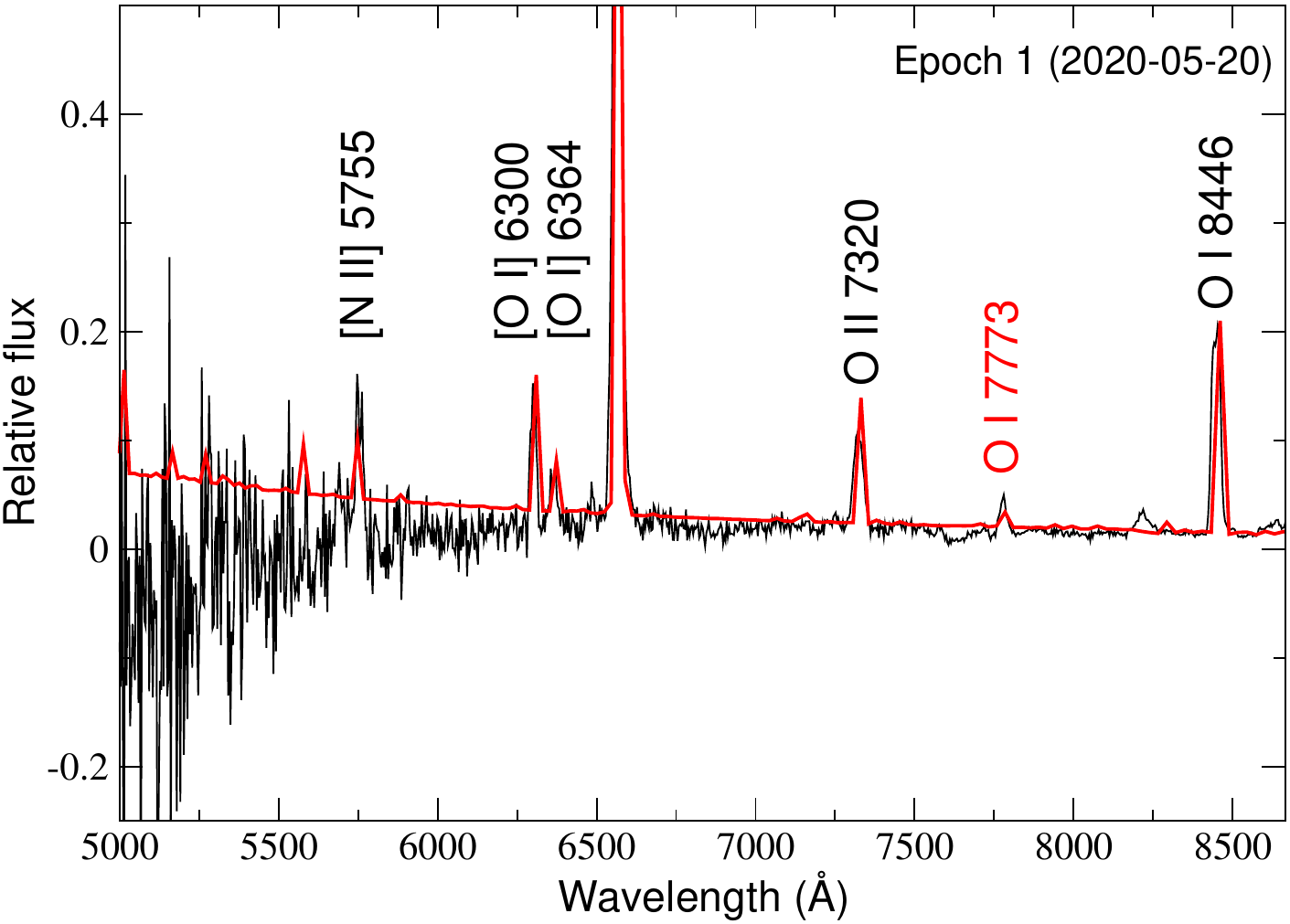}
  \caption{Observed and best-fit \textsc{cloudy} models for epoch 1 in the optical domain. The emission line of \ion{O}{i} 7773 {\AA} is labelled with red colour.}
  \label{fig:cloudy1}
\end{figure}
\begin{figure*}
    \centering
    \includegraphics[scale=0.55]{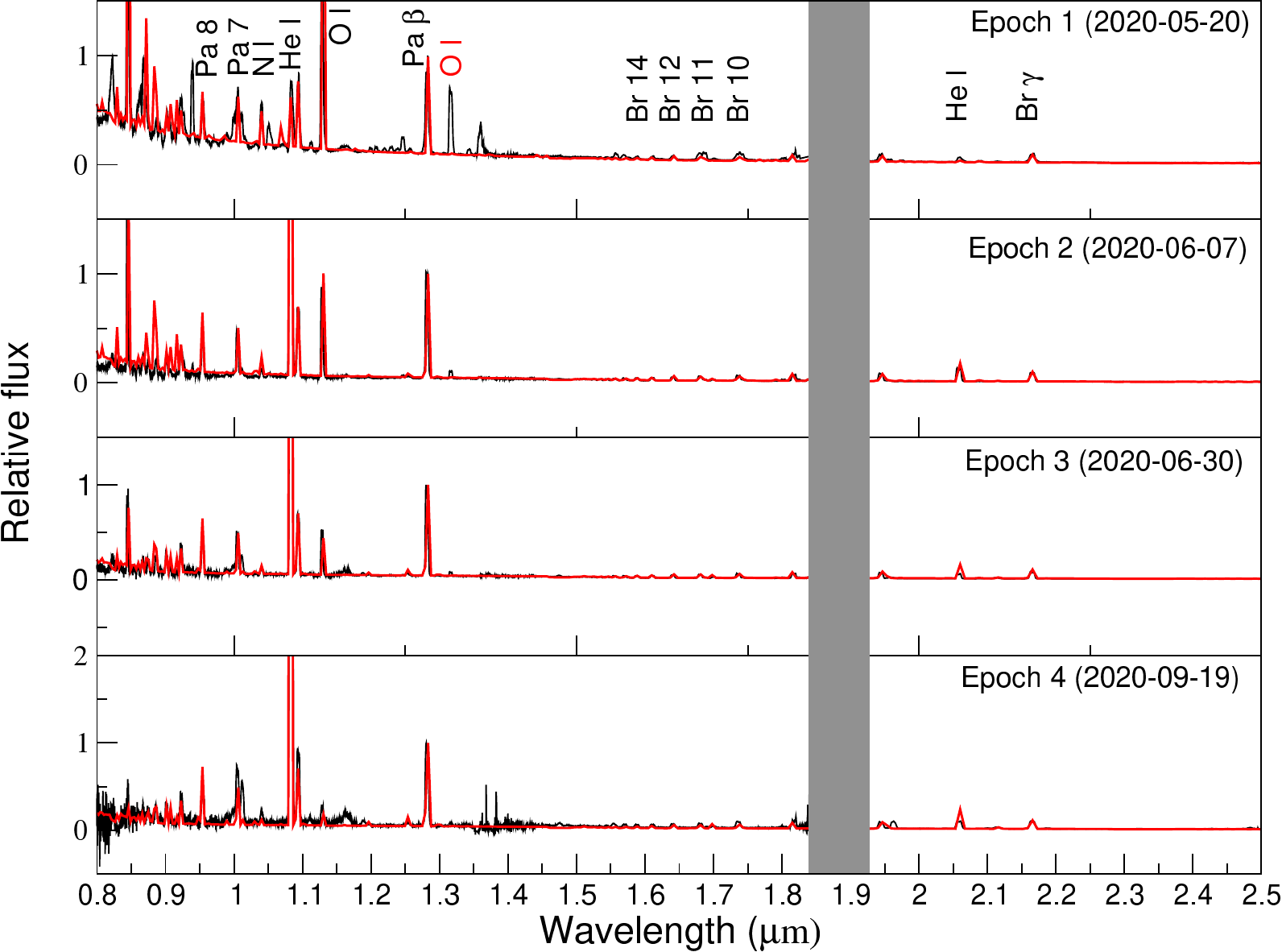}
    \caption{Observed and best-fit \textsc{cloudy} models for (a) epoch 1, (b) epoch 2, (c) epoch 3, and (d) epoch 4, NIR domain. The emission line of \ion{O}{i} 1.3164 $\mu$m is labelled with red colour.}
\label{fig:cloudy2}
\end{figure*}

The best-fit \textsc{cloudy} model parameters are shown in Table~\ref{tab:parameters}. Table~\ref{tab:chisquare} presents the relative flux of observed emission lines, modelled line ratios, and their related $\chi^2$ values. For $\chi^2$ calculations, only the lines that appear in both the \textsc{cloudy} model-generated and observed spectra were considered. Figure~\ref{fig:cloudy1} and~\ref{fig:cloudy2} depicts a comparison between model-generated best-fit spectra (represented by red solid lines) and observed spectra (represented by black lines) for epochs 1, 2, 3, and 4 in optical and NIR bands, respectively. Prominent emission lines are marked in the Figures. The line fluxes were estimated by interactive analysis using the \textit{splot} task of the \textit{onedpsec} package within \textsc{iraf} by fitting a Gaussian function. Flux ratios were determined for the \textsc{cloudy} model and observed data, with reference to \ion{H}{$\alpha$} and \ion{Pa}{$\beta$} in optical and NIR domains, respectively.

\begin{table*}
\caption{Best-fit \textsc{cloudy} model parameters.\label{tab:parameters}}
\setlength{\tabcolsep}{10pt}
\small
\begin{threeparttable}
\centering
\begin{tabular}{l c c c c c c c ccccc}
\hline
\hline
 Parameters           &	 Epoch 1   &	Epoch 2  	& Epoch 3 	& Epoch 4	 \\
                     &	  May 20   & June 09 &	June 30 	&   September 19\\ 

\hline
Temperature of central ionising WD  ($\times 10^4$K) & 3.09 & 3.98 & 5.37 & 6.31  \\
Luminosity of central ionising WD ($\times 10^{36}$erg s$^{-1}$) & 4.17 & 4.17 & 4.17 & 4.17 \\
Density of dense region of ejecta$^{c}$ ($\times 10^{8}$cm$^{-3})$  & 3.46 & 2.81 & 2.51 & 1.58 \\
Density of hot region of ejecta$^{c}$ ($\times 10^7$cm$^{-3})$ & 2.75 & 2.51 & 2.39 & 1.58 \\
dense to hot region covering factors$^{d}$ & 0.50/0.32 & 0.61/0.36 & 0.39/0.61 & 0.29/0.71 \\
$\alpha^{a}$	         &   -3     & -3  & -3	& -3	 	 \\
$r_{in}^{a}$  ($\times 10^{14}$ cm ) & 6.31 & 6.91 & 7.41 & 9.54 \\
$r_{out}^{a}$ ( $\times 10^{15}$cm)  & 1.73 & 1.86 & 2.04 & 2.57 \\
Filling factor$^a$	    &	  0.1	 & 0.1 & 0.1  	& 0.1   	 \\
$\beta^{a}$		          &	0.0	 & 0.0 & 0.0  	& 0.0  \\
He$^{b}$ &  0.35 $\pm$ 0.08 & 2.40 $\pm$ 0.20 & 2.50 $\pm$ 0.25 & 3.50 $\pm$ 0.35 \\
N$^{b}$  & 30$^{+5.00}_{-5.00}$ & 15$^{+6.00}_{-8.00}$ & 15$^{+6.00}_{-8.00}$ & 27$^{+5.00}_{-5.00}$ \\
O$^{b}$  &  9$^{+1.50}_{-2.00}$ & 7 $^{+1.05}_{-2.50}$  & 10$^{+1.55}_{-1.55}$ & 15$^{+3.00}_{-2.55}$  \\
Number of observed lines (n)  &  25  & 17 & 17 & 17\\
Number of free parameters (n$_{p}$) &  13   &   9     &  9        & 	9	 \\
Degrees of freedom ($\nu$)		  &  12  &	8     &	8    &   8	\\
Ejected mass ($\times 10^{-5} M_{\odot}$) & 6.71 & 6.82 & 3.54 & 5.30 \\
Total $\chi^{2}$ & 19.43 & 13.21 & 14.79 & 12.47	\\
$\chi^{2}_{red}$ & 1.61 & 1.65 & 1.84 & 1.55 \\
\hline
\end{tabular}
\begin{tablenotes}
\item (a) This was not a free parameter in the model.
\item (b) The log abundance by number relative to hydrogen, relative to solar values. The logs of solar abundances relative to hydrogen are He=-1.07, N=-4.17, and O=-3.31. All other elements not listed in the table were set to their solar values.
\item (c) Density values at the inner radius ($r_{in}$) of the ejecta.
\item (d) These numbers were only used to obtain the final spectrum by multiplying the spectrum of each component by their respective covering factors and then adding them.
\end{tablenotes}
\end{threeparttable}
\end{table*}

\subsection{Results}
\subsubsection{Central ionising source}\label{sec:centralsource}

The temperature of the central ionizing source during epoch 1 was determined to be $\sim 3.09 \times 10^4$ K, which increased to $\sim 6.31 \times 10^4$ K during epoch 4 (see Table~\ref{tab:parameters}), probably due to the collapse of the pseudo-photosphere onto the surface of WD as the nova evolved \citep{1976MNRAS.175..305B}. \citet{2022MNRAS.510.4265K} estimated the similar temperature of the central hot source in the range of 35000 - 54000 K following the empirical relation between the temperature of the hot source and the time to fall three magnitudes below the maximum \citep{1976MNRAS.175..305B,1989clno.conf...61B,2003AJEvans}. This estimate is based on the assumption that the nova pseudo-photosphere collapses at constant bolometric luminosity. Thus, the results obtained here are consistent with the physical picture of photosphere collapse \citep{2005MNRASEvans}.
\par
The luminosity of the central ionizing WD from the best-fit \textsc{cloudy} model was determined to be around $\sim 4.17 \times 10^{36}$ erg s$^{-1}$ for all epochs. The radius of the central ionising source using the best-fit \textsc{cloudy} model parameters were estimated to be $\sim 1.15$ R$_{\odot}$ for epoch 1, which decreased to $\sim 0.27$ R$_{\odot}$ for epoch 4. The best-fit model luminosity of the central ionising source gives $M_{bol} = -2.78$ by assuming $M_{bol} = 4.75$ for Sun \citep{2008A&AMunari}. 
\par 
In the constant bolometric luminosity phase, it is suggested \citep[e.g., see][]{1998PASP..110....3G,2008A&AMunari} that the total luminosity of the burning shell can be well described by the core mass-luminosity relationship proposed by \citet{1971AcAPaczy}:

\begin{equation}\label{eq:cml}
    \frac{L_{Pacz}}{L_{\odot}} = 6 \times 10^4 \Big( \frac{M_{WD}}{M_{\odot}} - 0.522 \Big).
\end{equation}

By substituting the luminosity value obtained from the best-fit \textsc{cloudy} model into the equation provided above, the resulting mass of the white dwarf is determined to be $M_{WD} = 0.54 M_{\odot}$. In a study by \citet{1992ApJLivio}, a relationship between the mass of a white dwarf and the decay time scale of a nova outburst, denoted as $t_3$, was presented. This relationship was established by utilising the white dwarf mass-radius relationship introduced by \citet{1972ApJNauenberg}, which can be expressed as:

\begin{equation}\label{eq:eqlivio1992}
t_3 = 51.3 \Big[\frac{M_{WD}}{M_{C}} \Big]^{-1} \Big[ \Big(\frac{M_{WD}}{M_{C}}\Big)^{-2/3} - \Big( \frac{M_{WD}}{M_{C}} \Big)^{2/3} \Big]^{3/2},
\end{equation}

where $M_{C} \simeq 1.4 ~M_{\odot}$ is the Chandrashekhar mass limit. We adopted an average value of $t_3 = 205$ from \citet{2022MNRAS.510.4265K} for V2891 Cyg. After substituting these values in equation~\ref{eq:eqlivio1992}, we obtained $M_{WD} \sim 0.55 M_{\odot}$. Estimating white dwarf masses is crucial in distinguishing between ONe (oxygen-neon) and CO (carbon-oxygen) type novae. The infrared (IR) evolution of CO novae, which results from explosions of WD of the CO type with a mass $M_{WD} < 1.2 M_{\odot}$, is primarily influenced by the onset and development of dust formation within the expanding material ejected from the explosion. In contrast, neon novae, alternatively known as ONe nova, originate from eruptions that transpire on white dwarfs of the ONe type, with a mass $M_{WD} > 1.2 M_{\odot}$. It is generally observed that these eruptions do not generate substantial quantities of dust \citep{1998PASP..110....3G}. Given the calculations of the mass of the WD using Equations~\ref{eq:cml} and~\ref{eq:eqlivio1992}, as well as the identification of the dust production epoch occurring during the latter stages of the V2891 Cyg nova, it is plausible to posit that the WD at its core is of the CO type.


\subsubsection{Ejecta density}
The ejecta density for two major components, namely Component 1 (ejecta with low hydrogen density) and Component 2 (ejecta with high hydrogen density), as determined by the best-fit \textsc{cloudy} model, is mentioned in Table~\ref{tab:parameters}. A detailed discussion on the third component, referred to as Component 3 (see Figure~\ref{fig:model}), which was added to the two-component model in order to generate \ion{O}{i} 7773 {\AA}, is presented in Section~\ref{sec:o1_7773}. 
\par
In our \textsc{cloudy} models, we noticed that the emission lines of \ion{He}{i} and [\ion{O}{ii}] 7320/7330 {\AA} were effectively generated by the low-density component (Component 1) of the two-component model, whereas the emission features of [\ion{O}{i}] 5577 {\AA}, [\ion{O}{i}] 6300 {\AA}, [\ion{O}{i}] 6364 {\AA}, and \ion{O}{i} $\lambda 8446$ {\AA} in the optical domain were reproduced by the high-density component ($n(H) > 10^8$ cm$^{-3}$, Component 2). The density of Component 1 and Component 2 at inner radius ($r_{in}$) of the ejecta was estimated to be around $2.75 \times 10^7$ cm$^{-3}$ and $3.46 \times 10^8$ cm$^{-3}$, respectively, during epoch 1. Subsequently, these values decreased to $1.58 \times 10^7$ cm$^{-3}$ and $1.58 \times 10^8$ cm$^{-3}$ for Component 1 and 2, respectively, by epoch 4. Furthermore, \ion{O}{i} $1.1286~\mu$m in the NIR domain was also reproduced by the high-density component. The most probable excitation mechanism for \ion{O}{i} 8446 {\AA} and \ion{O}{i} 1.1286 $\mu$m is through the Bowen mechanism of Ly $\beta$ fluorescence \citep{1947PASP...59..196B}. As nova ejecta expand, more Lyman $\beta$ photons can penetrate the outer regions of the ejecta, which are rich in neutral oxygen. This Lyman $\beta$ pumping process leads to an increase in the strength of \ion{O}{i} emission lines at $ 8446$ {\AA} and $ 1.1286~\mu$m.

\subsubsection{Chemical abundances}\label{subsec:abundance}

The derived elemental abundances from \textsc{cloudy} best-fit model parameters are shown in Table~\ref{tab:parameters}. Our best-fit models predicted enhanced values of He, N, and O with respect to their solar values. We used all prominent emission lines of oxygen to estimate the Oxygen abundance. We found that the O abundance is enhanced over its solar value by a factor of 9 in epoch 1, which is then increased to 15 during epoch 4. Other studies have also shown enhanced values of elemental abundances for the novae. \citet{Valandingham2005ApJ} reported He/H = 1.2 $\pm$ 0.2, N/H = 45 $\pm$ 11, and O/H = 22 $\pm$ 7 for nova V1974 Her. In a study by \citet{1997MNRASVanlandingham}, enhanced values of N/H $\sim$ 135 and O/H $\pm$ 24 for Nova V693 Coronae Austrinae 1981 was reported. For nova GQ MUS, \citet{1996A&AMorisset} found He/H = 2.7, N/H = 316, O/H = 36. In a recent study on the Nova V1674 Her, \citet{2024MNRASHabtie} found He/H $\sim$ 2.5 - 2.9, N/H $\sim$ 30 - 50, O/H $\sim$ 5 - 10. Observations have revealed that the novae ejecta are typically heavily enriched in CNO elements \citep{1998PASP..110....3G,2012ApJHelton,2021JApADas}. This enrichment is generally explained by considering the possibility of mixing accreted material with the core envelope of the underlying WD before the final stages of TNR \citep[see, e.g.][etc.]{1972ApJStarrfield,1994ApJLivio,2012BASIJose}. As a result, the expelled gases during an outburst are the combination of a mixture of WD and accreted material from the secondary that has been processed by hot-hydrogen burning during TNR \citep{2020ApJStarrfield}. Among CNO elements, Nitrogen is usually the most abundant element in the nova ejecta as the accreted material is mixed with the white dwarf material and processed through the incomplete CNO cycle \citep{1986ApJTruran} that changes the relative abundance of the CNO elements. From our best-fit models, we found a substantial enhancement of the nitrogen in V2891 Cyg ejecta, approximately 30 times higher than its solar value. In addition to the enhanced CNO heavy elements, the enrichment of He is also observed in novae ejecta (e.g., GQ Mus, HR Del, RR Pic, etc.), implying that the He shell and the accreted material may undergo a mixing process \citep[see e.g.][]{1986ApJTruran}. The high abundance of Oxygen in our best-fit model might also indicate that the underlying WD in V2891 Cyg might be CO type \citep{2012ApJHelton,2017A&ABalman}. For a detailed discussion on elemental abundance in novae, please see \citet[][and references therein]{2012ApJHelton,2021JApADas}. A compiled list of derived elemental abundances for novae can be found in \citet{1998PASP..110....3G,josé_shore_2008}.
\par

\subsubsection{Origins of O I 7773 {\AA} emission}\label{sec:o1_7773}

The \ion{O}{i} 7773 {\AA} emission line in the spectral evolution is notable for its time-varying P-Cygni profile, suggesting the possibility of multiple mass ejections \citep{2022MNRAS.510.4265K}. It should be noted that the two-component phenomenological models were unable to generate the emission line of \ion{O}{i} 7773 {\AA}. We found that a third dense and cool component of high oxygen abundance was needed to generate the \ion{O}{i} 7773 {\AA} line (see Figure~\ref{fig:model}). 
\par 
As discussed in Section~\ref{subsec:7773and13164}, the \ion{O}{i} 7773 {\AA} line showed the time-varying P-Cygni profile, coinciding with the re-brightening event observed in the nova, in contrast to other Oxygen lines. In addition, \citet{2012AJWilliams} studied the spectral development of \ion{Fe}{ii} and He/N type novae and argued that it is not very common for the \ion{O}{i} 7773 {\AA} line to be very strong and comparable to the \ion{O}{i} 8446 {\AA} line in the optical spectrum. \citet{Kastner1995ApJ} in their theoretical models proposed that at higher densities, collisional transfer between the \ion{O}{i} triplet and quintet states will significantly enhance \ion{O}{i} 7773 {\AA} emission line flux. Considering the fact that the best-fit ejecta density of the two-component model was not sufficient to generate this line and the presence of a time-varying P-Cygni profile in this line, which was absent in other \ion{O}{i} lines, we hypothesized that this line might have originated from separate dense clumps within the ejecta, supporting the idea of multiple mass ejections of dense clumps with high O abundance. 
\par
In our model, we adopted an expansion velocity of 300 km s$^{-1}$ for this third component, which is estimated from the blue dip of the P-Cygni profile (which was found to be in the range of 300 - 750 km s$^{-1}$ \citep{2022MNRAS.510.4265K}), and estimated the distance to be around, $R_{\text{OI}} \sim 10^{14.8}$ cm for epoch 1. Therefore, we set the $R_{in} = 10^{14.75}$ cm, and the thickness of this shell is around 0.1 $R_{\text{OI}}$ \citep{2012ApJHelton} for this third component (see Figure~\ref{fig:model}). To limit degrees of freedom, we set the abundances of other elements at their solar values from \citet{2010Ap&SS.328..179G}, focusing solely on generating the \ion{O}{i} 7773 {\AA} line. The temperature of this third component was estimated to be in the range of $\sim$ 7049 - 7413 K, with an oxygen abundance of approximately 28 times its solar value. We found that the covering factor of this third component while generating the final spectrum was 0.11. We adjusted the covering factors from all three components in the final spectrum so that their total sum is less than or equal to one. The final spectrum is shown in Figure~\ref{fig:cloudy1}.
\par
It is well known that novae have complicated density structures, and clumps are embedded within the ejecta. Since details about the clumps are not known, we simply added a third component in the model to check the origin of \ion{O}{i} 7773 {\AA} emission from a region of very high ejecta density ($\sim 10^{11}$ cm$^{-3}$). These values are aligned with the temperature reported by \citet{Kastner1995ApJ} to generate an \ion{O}{i} line in classical novae. Also, see section~\ref{subsec-OxygenLines} for a detailed discussion on \ion{O}{i} 7773 {\AA} emission line in novae. Hence, it is plausible to argue that there were indeed dense clumps with high O abundance present within the ejecta, serving as the source of emission for the \ion{O}{i} 7773 {\AA} line in V2891 Cyg. This hypothesis is further supported by the observational fact that in contrast to the \ion{O}{i} 7773 {\AA} line, the remaining \ion{O}{i} lines do not exhibit time-varying P-Cygni profiles, leading to a possibility that \ion{O}{i} 7773 {\AA} originates from a different region of the ejecta.


\subsubsection{Mass in the shell}
The hydrogen mass contained in model ejecta could be estimated using the following relation from \cite{2001MNRASSchwarz},
\begin{equation}\label{eq:mass}
   M_{\text{shell}} = \mu \int n(r)f(r) dV(r)
\end{equation}
where $n(r)$, and $f(r)$ are the hydrogen number density (cm$^{-3}$), and filling factor, respectively. $\mu$ is the mean atomic mass that represents the total mass fraction of all best-fit elemental abundances relative to hydrogen within the ejecta, which can be written as;
\begin{equation}\label{eq:mu}
    \mu = \sum_n (1 + A_n) m_p,
\end{equation}
where $m_p$ is the mass of the proton. Using Equation~\ref{eq:nh}, \ref{eq:filling}, and \ref{eq:mu}, equation~\ref{eq:mass} can be written to estimate the total ejecta mass:
\begin{equation}\label{eq:finalmass}
M_{\text{shell}} = n(r_{\text{in}})f(r_{\text{in}}) \sum_n (1 + A_n) m_p \int_{r_{\text{in}}} ^{r_{\text{out}}} (r/r_{\text{in}})^{\alpha + \beta} 4\pi r^2 dr.
\end{equation}
The density, filling factor, the values of $\alpha$, and $\beta$ of the shell are set by the best-fit model density (Table~\ref{tab:parameters}). For the present study, the \textsc{cloudy} model assumed two components and an additional third component for \ion{O}{i} 7773 {\AA}. It is important to note that the third component was only included in epoch 1, whereas all later epochs used two-component models that solely utilised NIR spectra for modelling. The total shell mass could be estimated by multiplying the shell mass from all components with their respective contributing percentage and then adding them. The \textsc{cloudy} best-fitting model for epochs 1, 2, 3, and 4, predicted the ejected hydrogen shell mass of $\sim 6.71 \times 10^{-5} M_{\odot}$, $6.82 \times 10^{-5} M_{\odot}$, $3.54 \times 10^{-5} M_{\odot}$, $5.30 \times 10^{-5} M_{\odot}$, respectively. The average estimated mass is around $5.59 \times 10^{-5} M_{\odot}$.
While estimating the mass of the ejecta, we assumed $\mu = 1$. However, analysis of our \textsc{cloudy} best-fit model parameters reveals an overabundance of helium. Accounting for this overabundance, the estimated ejected mass would change by a factor of 1.1 for epoch 1, 1.8 for epochs 2 and 3, and 2.1 for epoch 4. It's important to note that our use of $\mu=1$ as the mean atomic mass in Equation~\ref{eq:finalmass} may have resulted in our reported numbers potentially representing a lower limit on the total ejected mass. These estimated masses are consistent with those found by \citet{2022MNRAS.510.4265K}, who employed Case B analysis of hydrogen emission lines and estimated a value of $8.60 \pm 1.73 \times 10^{-5} M_{\odot}$. The mass of the Component 3 was estimated to be around $1.95 \times 10^{-5} M_{\odot}$. 
\par 
\citet{Shore1993AJ} proposed that the total ejecta mass of nova with He overabundance, which adheres to a linear velocity law and an inverse cube density law, may be approximated using the following expression,

\begin{equation}\label{eq:shore1993}
    M_{ej} \sim Y^{-1/2} 10^{-4} M_{\odot}
\end{equation}
where $Y$ represents the enhancement factor for the helium abundance. By using the value of $Y$ from the best-fit \textsc{cloudy} model parameters (Table~\ref{tab:parameters}), the mass of the ejected shell for epochs 1, 2, 3, and 4 are found to be: $2.97 \times 10^{-6} M_{\odot}$, $2.55 \times 10^{-5} M_{\odot}$, $3.06 \times 10^{-5} M_{\odot}$, $3.2 \times 10^{-5} M_{\odot}$, respectively. It should be noted that Equation~\ref{eq:shore1993} by \citet{Shore1993AJ} was used for Nova Cygni 1992, a neon nova, whereas V2891 Cyg is most likely a CO type nova. However, the ejecta mass estimated using Equation~\ref{eq:shore1993} for V2891 Cyg is consistent with the ejecta mass estimation by \citet{2022MNRAS.510.4265K}.

\subsubsection{On the utility of photo-ionization models in explaining O I 7773 {\AA} and 1.3614 $\mu$m emissions}
\par
The phenomenological photoionization model presented above has successfully reproduced the observed properties of V2891 Cyg to a great extent. The derived values are consistent with the typical parameters associated with a nova outburst. As emphasised in section~\ref{subsec:7773and13164}, two of the neutral oxygen lines  \ion{O}{i} 7773 {\AA} and \ion{O}{i} 1.3614 $\mu$m, showed an evolution pattern distinct from other \ion{O}{i} lines. We attempted to reproduce these observed emission lines using two-component photoionization models, which have been successfully used for other nova outbursts (see Section~\ref{sec:cloudy}). However, these models failed to produce both of these lines for V2891 Cyg. We reproduced the \ion{O}{i} 7773 {\AA} line by incorporating a dense third component to the model with only a fraction of total ejecta mass. This is consistent with the hypothesis proposed by \citet{2022MNRAS.510.4265K} that the nova had undergone multiple mass ejections during its early outburst phase, signified by multiple peaks in its light curve and the corresponding appearance of P-Cygni profile in \ion{O}{i} 7773 {\AA}. It is, thus, supported by our \textsc{cloudy} modelling that oxygen embedded in these ejected globules of high density is responsible for the appearance/disappearance of the P-Cygni profile in \ion{O}{i} 7773 {\AA} line.
\par
Even after considering a third component, our photoionization phenomenological models could not generate the \ion{O}{i} 1.3614 $\mu$m line in the NIR spectra of V2891 Cyg. The flux variation in this line coincided with a brief epoch of dust formation and coronal emission, suggesting a potential connection between this emission and the dust formation process in the nova. \citet{Derdzinski2017MNRAS} has presented a scenario where a thin-dense region is created between forward and reverse shocks where dust could be formed. Due to the fewer NIR spectra and a modest amount of dust \citep{2022MNRAS.510.4265K}, it was not possible to explore the composition of dust in this case. We, thus, tried to reproduce the scenario (a phenomenological model) wherein \ion{O}{i} 1.3614 $\mu$m can originate within this thin, dense shell using collisional ionization models of \textsc{cloudy}. The spectral synthesis code \textsc{cloudy} is proficient at simulating both photoionization models and collisional ionisation models. It has the ability to accurately simulate the environments in which ionisation processes occur as a result of high-energy collisions. The collisional ionization rate coefficients for collisional ionization models in \textsc{cloudy} are computed from \citet{1997ADNDTVoronov}. 
\begin{landscape}
\begin{table}
\setlength{\tabcolsep}{4pt}
\renewcommand{\arraystretch}{1.2}
\begin{center}
\caption{Observed and best-fit \textsc{cloudy} model line flux ratios for V2891 Cyg.  Observed and modelled fluxes are normalised by \ion{H}{$\alpha$} and \ion{Pa}{$\beta$} in optical and NIR bands, respectively. \label{tab:chisquare}}
\begin{tabular}{l l c c c c c c c c c c c c}
\hline\noalign{\smallskip}
\hline
 &            &          &      Epoch 1 &                &              &  Epoch 2 &            &       &       Epoch 3  &                    &       &  Epoch 4  &  \\
$\lambda$  & Line ID  & Observed Flux     &  Model Flux    & $\chi^{2}$ & Observed Flux &  Model Flux     & $\chi^{2}$ & Observed Flux  &  Model Flux     & $\chi^{2}$ & Observed Flux  & Model Flux     & $\chi^{2}$\\
\hline\noalign{\smallskip}
 \multicolumn{14}{c}{Optical}\\\hline
5577 {\AA} & [\ion{O}{i}]  & 2.90E-01 & 2.50E-01 & 1.60E-01 & ... & ... & ... & ... & ... & ... & ... & ... & ... \\
5755 {\AA} & [N II] & 1.38E-01 & 1.20E-01 & 3.24E-02 & ... & ... & ... & ... & ... & ... & ... & ... & ... \\
6300 {\AA} & [\ion{O}{i}] & 1.00E-01 & 1.01E-01 & 1.00E-04 & ... & ... & ... & ... & ... & ... & ... & ... & ... \\
6364 {\AA} & [\ion{O}{i}]  & 5.70E-02 & 7.00E-02 & 1.60E-02 & ... & ... & ... & ... & ... & ... & ... & ... & ...\\
6563 {\AA} & H $\alpha$& 1.00E+00 & 1.00E+00 & 0.00E+00 & ... & ... & ... & ... & ... & ... & ... & ... & ...\\
7320 {\AA} & \ion{O}{ii} & 1.10E-01 & 1.80E-01 & 4.90E-01 & ... & ... & ... & ... & ... & ... & ... & ... & ...\\
7773 {\AA} & \ion{O}{i} & 3.38E-02 & 1.90E-02 & 2.19E-02 & ... & ... & ... & ... & ... & ... & ... & ... & ...\\
8446 {\AA} & \ion{O}{i} & 2.15E-01 & 2.01E-01 & 1.96E-02 & ... & ... & ... & ... & ... & ... & ... & ... & ...\\
 \hline 
 \multicolumn{14}{c}{NIR}\\\hline
0.9546 $\mu$m & Pa 8 & 3.68E-01 & 3.51E-01 & 2.89E-02 & 2.67E-01 & 1.98E-01 & 4.76E-01&3.41E-01 & 2.88E-01 & 2.81E-01 & 4.01E-01 & 3.83E-01 & 3.24E-02 \\
1.0047 $\mu$m & Pa 7 & 3.65E-01 & 3.55E-01 & 9.22E-03 & 2.87E-01 & 3.11E-01 & 5.76E-02&2.91E-01 & 1.68E-01 & 1.51E+00 & 4.37E-01 & 4.11E-01 & 6.76E-02 \\
1.0398 $\mu$m & N I & 3.78E-01 & 3.50E-01 & 7.73E-02 & 7.10E-02 & 8.20E-02 & 1.21E-02&6.80E-02 & 6.00E-02 & 6.40E-03 & 1.44E-01 & 1.62E-01 & 3.24E-02 \\
1.083 $\mu$m & He I & 7.80E-01 & 6.40E-01 & 1.96E+00 & 5.98E+00 & 6.15E+00 & 2.89E+00&9.34E+00 & 9.25E+00 & 8.10E-01 & 1.37E+01 & 1.40E+01 & 7.84E+00 \\
1.0938 $\mu$m & Pa $\gamma$ & 7.60E-01 & 7.10E-01 & 2.50E-01 & 5.20E-01 & 4.14E-01 & 1.12E+00&5.50E-01 & 5.20E-01 & 9.00E-02 & 7.80E-01 & 6.30E-01 & 2.25E+00 \\
1.1126 $\mu$m & Fe II & 1.00E-01 & 1.10E-01 & 1.00E-02 & 1.80E-02 & 7.00E-03 & 1.21E-02&2.50E-02 & 2.00E-02 & 2.50E-03 & 1.60E-02 & 9.00E-03 & 4.90E-03 \\
1.1286 $\mu$m & \ion{O}{i} & 1.51E+00 & 1.91E+00 & 1.63E+01 & 6.80E-01 & 8.80E-01 & 4.00E+00&4.03E-01 & 3.81E-01 & 4.84E-02 & 1.90E-01 & 1.30E-01 & 3.60E-01 \\
1.2527 $\mu$m & He I & 5.60E-02 & 5.10E-02 & 2.50E-03 & 1.00E-01 & 1.10E-01 & 1.00E-02&4.50E-02 & 4.91E-02 & 1.68E-03 & 4.60E-02 & 5.00E-02 & 1.60E-03 \\
1.2818 $\mu$m & Pa $\beta$ & 1.00E+00 & 1.00E+00 & 0.00E+00 & 1.00E+00 & 1.00E+00 & 0.00E+00&1.00E+00 & 1.00E+00 & 0.00E+00 & 1.00E+00 & 1.00E+00 & 0.00E+00 \\
1.5439 $\mu$m & Br 17 & 3.00E-02 & 2.65E-02 & 1.22E-03 & 1.83E-02 & 2.10E-02 & 7.29E-04&3.29E-02 & 3.30E-02 & 1.00E-06 & 3.20E-02 & 4.00E-02 & 6.40E-03 \\
1.5881 $\mu$m & Br 14 & 3.10E-02 & 2.90E-02 & 4.00E-04 & 3.44E-02 & 3.60E-02 & 2.56E-04&3.01E-02 & 3.00E-02 & 1.00E-06 & 3.40E-02 & 4.40E-02 & 1.00E-02 \\
1.6109 $\mu$m & Br 13 & 3.70E-02 & 3.20E-02 & 2.50E-03 & 3.79E-02 & 4.00E-02 & 4.41E-04&3.81E-02 & 3.85E-01 & 1.20E+01 & 5.30E-02 & 5.80E-02 & 2.50E-03 \\
1.6407 $\mu$m & Br 12 & 4.80E-02 & 4.00E-02 & 6.40E-03 & 4.88E-02 & 5.10E-02 & 4.84E-04&5.36E-02 & 5.40E-02 & 1.60E-05 & 5.50E-02 & 6.10E-02 & 3.60E-03 \\
1.6806 $\mu$m & Br 11 & 8.80E-02 & 8.20E-02 & 3.60E-03 & 1.15E-01 & 1.84E-01 & 4.76E-01&1.55E-01 & 1.51E-01 & 1.60E-03 & 5.90E-02 & 6.20E-02 & 9.00E-04 \\
1.7362 $\mu$m & Br 10 & 1.68E-01 & 1.70E-01 & 4.00E-04 & 1.28E-01 & 1.10E-01 & 3.10E-02&6.50E-02 & 6.30E-02 & 4.00E-04 & 6.90E-02 & 6.90E-02 & 0.00E+00 \\
2.0585 $\mu$m & He I & 8.80E-02 & 8.00E-02 & 6.40E-03 & 2.38E-01 & 4.41E-01 & 4.12E+00&9.20E-02 & 1.00E-01 & 6.40E-03 & 1.74E-01 & 3.10E-01 & 1.85E+00 \\
2.1655 $\mu$m & Br $\gamma$ & 1.60E-01 & 1.57E-01 & 9.00E-04 & 1.42E-01 & 1.50E-01 & 6.40E-03&1.37E-01 & 1.36E-01 & 1.00E-04 & 1.62E-01 & 1.50E-01 & 1.44E-02 \\
\hline
Total &  & & & 19.43  & & &  13.21  & & & 14.79 & & & 12.47\\
\noalign{\smallskip}\hline
\end{tabular}
\end{center}
\end{table}
\end{landscape}

As we shall discuss, the photo-ionization again failed to re-produced the observed flux of this line; therefore, we focused on the use of collisional ionization models in \textsc{cloudy} as a potential alternate approach for reproducing \ion{O}{i} 1.3614 $\mu$m emission in V2891 Cyg.

\begin{figure*}
    \centering
    \includegraphics[width=0.6\linewidth]{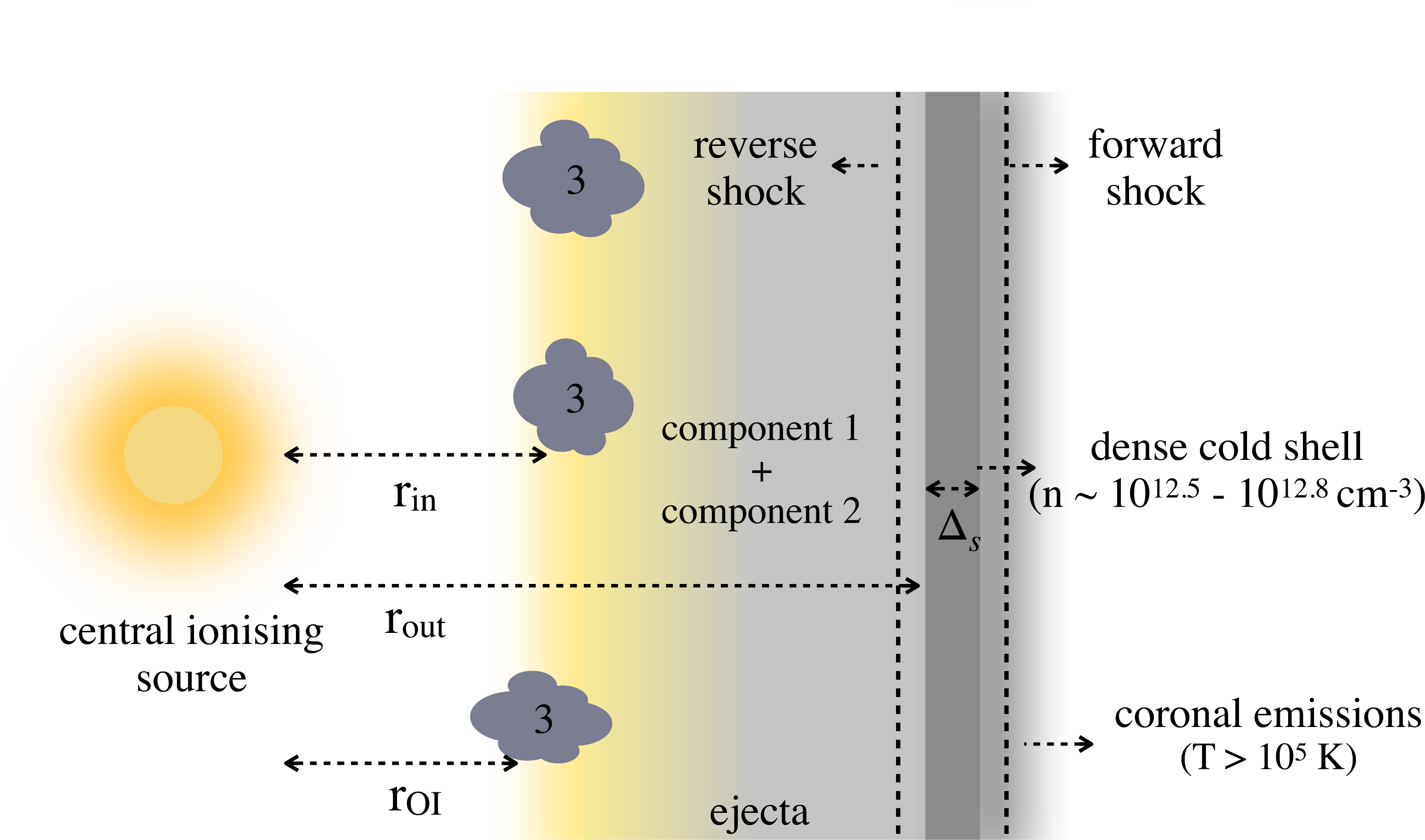}
    \caption{Schematic diagram of the phenomenological model used for \textsc{cloudy} photoionization and collisional ionization models. Please see Section~\ref{sec:cloudy} and~\ref{sec:OIlines_shock} for detailed discussion. The figure is not to scale.}
    \label{fig:model}
\end{figure*}


\section{A Collisional Ionization model for the dust forming layer: the origin of O I 1.3164 micron emission line}\label{sec:OIlines_shock}

As discussed above, our \textsc{cloudy} photoionization phenomenological models for the nebular phase could not produce \ion{O}{i} 1.3164 $\mu$m feature (refer to Figures~\ref{fig:cloudy2} in Section~\ref{sec:cloudy}). In this section, we explored the collisional ionization models of \textsc{cloudy} to generate this observed emission line in the spectra. \citet{Metzger2014MNRAS} studied the internal shocks in novae produced by the interaction between a fast outflow of a nova and a dense circumstellar shell and proposed that these shocks become radiative when the density of the ejecta reach substantial levels. The gas behind these shocks cools rapidly and piles up in a central cold shell sandwiched between the forward and reverse shocks \citep[see Figure 1 in][]{Metzger2014MNRAS}. In their theoretical shock models for novae, \citet{Metzger2014MNRAS} also made an estimation of the temperature of the thin shell created by the shock, which was found to be approximately T $\sim 10^4$ K. Following these results by \citet{Metzger2014MNRAS}, \citet{Derdzinski2017MNRAS} proposed a dust shock model where they argued that this dense shell provides an ideal and shielded site for the nucleation of dust grains. They simulated dust shock models in novae and demonstrated that the shock raises the temperature of the gas to an exceedingly high level while also compressing it.
\par 
\citet{2022MNRAS.510.4265K} also proposed the possibility of multiple mass ejections in V2891 Cyg, perhaps leading to the formation of internal shocks. These shocks may facilitate dust formation in the thin, dense layer formed by the shocks within the ejecta. Building on these arguments, we considered a geometry for the shock-induced dust model for nova as proposed by \citet{Derdzinski2017MNRAS}. We assumed that the internal shocks were formed in the V2891 Cyg ejecta and that the post-shock gas could cool efficiently and create a thin and dense shell between the forward and reverse shocks in the ejecta of V2891 Cyg. Following the argument by \citet{Derdzinski2017MNRAS} that this neutral region is precisely where dust is expected to form, we considered this thin shell to be located in the outer region of the ejecta at $r_{out}$ from the central ionising source. We implemented this in a simple phenomenological model to explore whether the emission of \ion{O}{i} 1.3164 $\mu$m originated from this dense, thin region. A schematic diagram illustrating the phenomenology is provided in Figure~\ref{fig:model}.
\par
\begin{figure*}
  \centering
  \begin{minipage}[b]{0.45\textwidth}
    \centering
    \includegraphics[width=\textwidth]{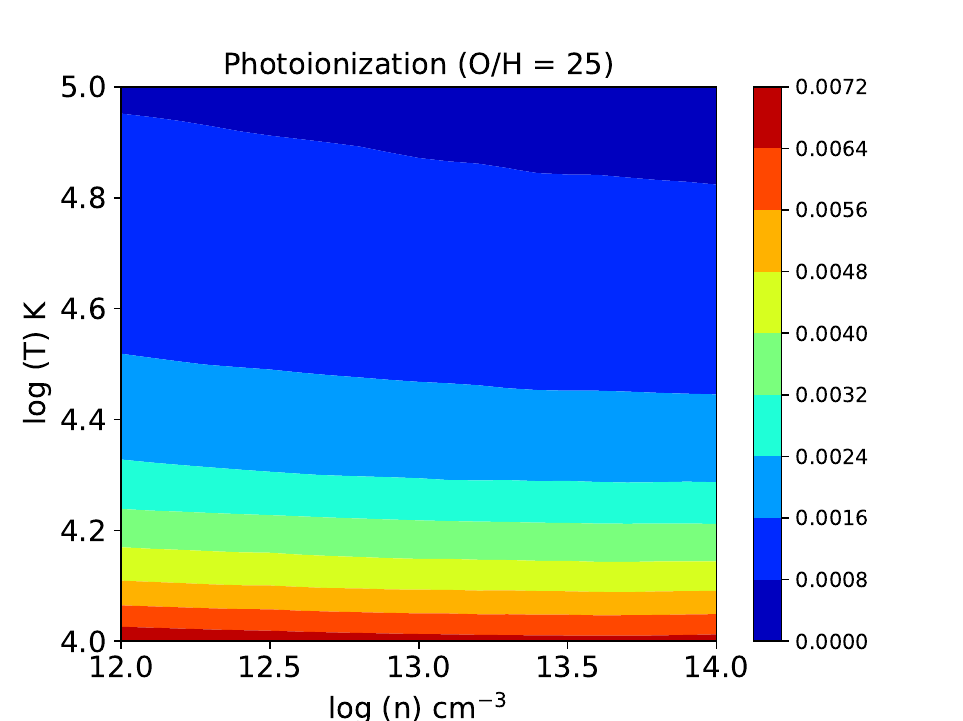}
    \label{fig:fig1}
  \end{minipage}
  \quad
  \begin{minipage}[b]{0.45\textwidth}
    \centering
    \includegraphics[width=\textwidth]{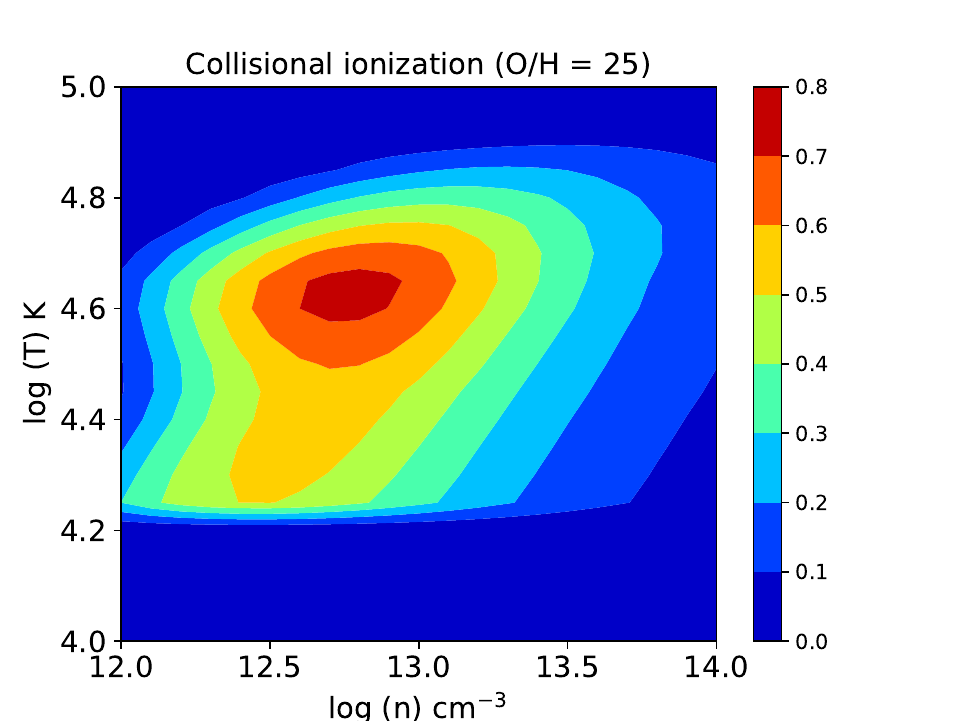}
    \label{fig:fig2}
  \end{minipage}
  \vspace{0.3cm} 
  \begin{minipage}[b]{0.45\textwidth}
    \centering
    \includegraphics[width=\textwidth]{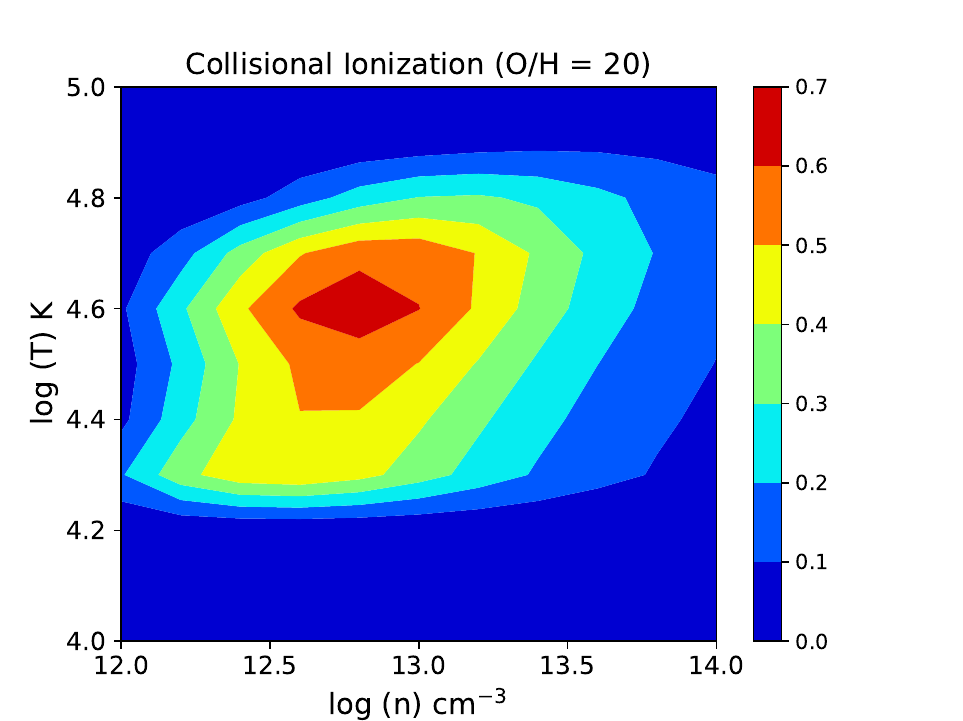}
    \label{fig:fig3}
  \end{minipage}
  \quad
  \begin{minipage}[b]{0.45\textwidth}
    \centering
    \includegraphics[width=\textwidth]{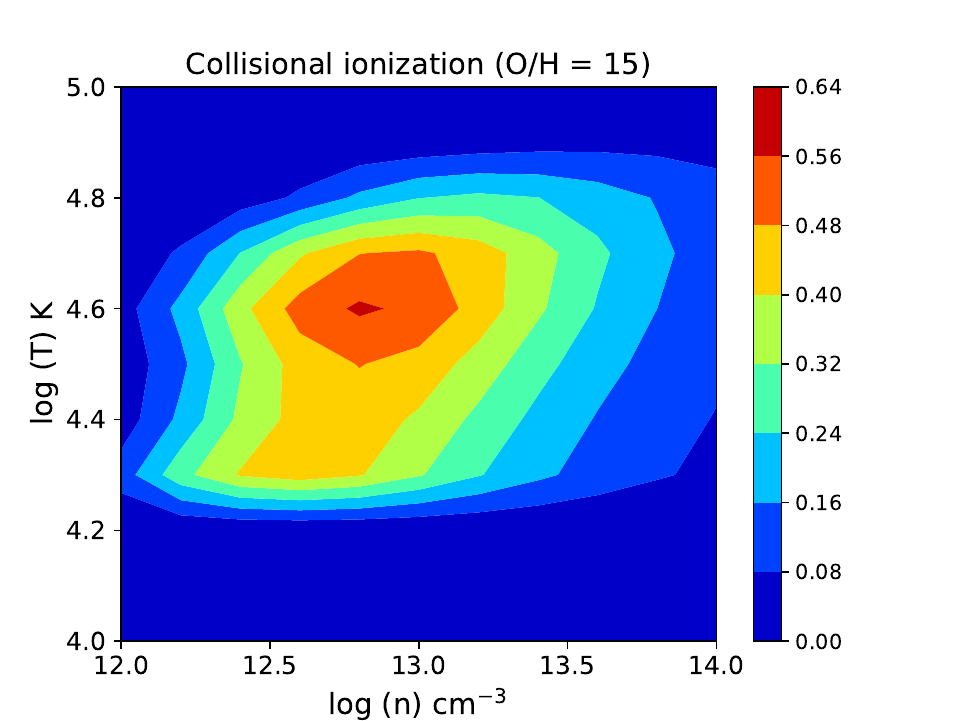}
    \label{fig:fig4}
  \end{minipage}
\caption{The contour plot for the variation of flux ratio of \ion{O}{i} 1.3164 / Pa $\beta$ with the hydrogen density (n) and temperature (T). (a) \textsc{cloudy} models with photoionization case, with O/H = 25, and the collisional ionization models for O/H = 15, 20, and 25 are shown in (b), (c), and (d), respectively.}
\label{fig:contourOI13164}
\end{figure*}

 
The thickness ($\Delta_s \sim 10^{10.8}$cm) and filling factor ($f(r) = 0.5$) of this thin and dense shell was adopted from values proposed in \citet{Derdzinski2017MNRAS}. We ran a grid of several models by varying the temperature of this thin shell in the range from $10^4 - 10^5$ K and the ejecta density in the range from $10^{12.0} - 10^{14.0}$ cm$^{-3}$. The range of these parameters was adapted from the previously published literature \citep{Metzger2014MNRAS,Derdzinski2017MNRAS}. We employed \textsc{cloudy} simulations to generate the \ion{O}{i} 1.3164 $\mu$m line by considering both photoionization and pure collisional ionization scenarios. The contour plots of the simulated line ratio of \ion{O}{i} 1.3164 $\mu$m / Pa $\beta$ for different oxygen abundance are shown in Figure~\ref{fig:contourOI13164}. These simulations were conducted for the observed NIR data for all four epochs (see Table~\ref{tab:1.3164}). The observed \ion{O}{i} 1.3164 $\mu$m / Pa $\beta$ ratio during epochs 1, 2, 3, and 4 was $\sim$ 0.73, 0.06, 0.04, and 0.03, respectively. Initially, we considered the photoionization scenario in \textsc{cloudy} models and observed that, even with a very high oxygen abundance (O/H = 25), the models failed to replicate the observed flux strength of the \ion{O}{i} 1.3164 $\mu$m / Pa $\beta$ line ratio (see Figure~\ref{fig:contourOI13164}(a)). Thereafter, we removed the central photoionizing source in pure collisional ionisation \textsc{cloudy} simulations and run a grid of models by varying the temperature and density of this thin, dense shell. Notably, these models demonstrated an increase in the flux ratio of the \ion{O}{i} 1.3164 $\mu$m / Pa $\beta$ emission lines (see Figure~\ref{fig:contourOI13164}). The peak of this increase was identified around the temperature of $10^{4.5} - 10^{4.6}$ K, which was consistent with the observed ratio of $\sim$ 0.7. Furthermore, the density of this thin shell was estimated to be approximately in the range $10^{12.5} - 10^{12.8}$ cm$^{-3}$ - roughly four orders of magnitude higher than the density of the pre-shocked region in our model. Only the oxygen abundance of this shell was varied in these simulation runs, maintaining other abundances at their solar values from \citet{2010Ap&SS.328..179G}, as our focus was solely on reproducing the \ion{O}{i} 1.3164 $\mu$m flux. Our analysis showed that, for epoch 1, our phenomenological model successfully matched the observed flux ratio with a high oxygen abundance of O/H = 25. Subsequently, the observed flux ratio decreased in successive epochs, indicating a decline in the oxygen abundance from epochs 1 to 4. To enhance the computational efficiency of the \textsc{cloudy} grid simulations, we have decided to narrow the investigation of oxygen abundance to a range of 5 to 25, with a step size of 5, which was found to be sufficient to reproduce the range of observed flux ratio. Our simulations have shown a higher amount of oxygen in the proposed thin shell, which supports our previous discussion about the possibility of oxygen being distributed unevenly in the material ejected by V2891 Cyg (see section~\ref{sec:o1_7773}). The derived enhanced oxygen abundance is corroborated by several additional research, as explained in the previous section~\ref{subsec:abundance}. Previous studies have also identified cases in which oxygen-rich material is dispersed throughout various regions of the nova's ejected material \citep{2012ApJHelton}. It should be noted that the two-component photoionization \textsc{cloudy} model (in Section~\ref{sec:cloudy}) estimated abundance in the inner part of the ejecta, revealing lower values than those estimated for this thin shell and component-3. The collisional ionization model of \textsc{cloudy} in Section 4 operates independently of the model used in Section~\ref{sec:cloudy}; however, they still contribute to a unified understanding of the distribution of \ion{O}{i} in the nova ejecta. In this scenario, the strong internal shocks lead to rapid cooling of the post-shock material, causing an accumulation of material with high O abundance in a thin and dense shell between the forward and reverse shocks. The shell's estimated density is roughly $10^{12.5} - 10^{12.8}$ cm$^{-3}$, which aligns with the results of \citet{Derdzinski2017MNRAS} who stated that thin, dense layers can have densities of up to around $10^{13}$ cm$^{-3}$.

Our collisional ionization phenomenological models of \textsc{cloudy} provide additional support, indicating that this thin region was indeed formed within the ejecta of V2891 Cyg, and most likely the emission of \ion{O}{i} 1.3164 $\mu$m was originated from this region. We also observed a noticeable reduction in the observed flux of this emission line during the subsequent epochs of dust formation. This decline was effectively replicated by the decreased oxygen abundance and filling factor in our phenomenological models, as given in Table~\ref{tab:1.3164}.

\begin{table}
\caption{Comparison between observed and modelled line fluxes of \ion{O}{i} 1.3164 $\mu$m / Pa $\beta$.\label{tab:1.3164}}
\setlength{\tabcolsep}{3pt}
\small
\begin{threeparttable}
\centering
\begin{tabular}{l c c c c c c c ccccc}
\hline
\hline
 Date    & O/H & Modelled & Observed &  $M/O$ & $r.m.s^{b}$ \\
        &     & flux$^{a}$ ($M$) & flux ($O$) &  &  \\        
\hline
2020-05-20 & 25.0 & $7.294 \times 10^{-1}$ & $7.217 \times 10^{-1}$ & 1.0106 & 0.0106\\
2020-06-07 & 13.2 & $6.597 \times 10^{-2}$ & $6.563 \times 10^{-2}$ & 1.0091 & 0.0091\\
2020-06-30 & 10.55 & $4.085 \times 10^{-2}$ & $4.058 \times 10^{-2}$ & 1.0067 & 0.0067\\
2020-09-19 & 9.6 & $3.337 \times 10^{-2}$ & $3.336 \times 10^{-2}$ & 1.0002 & 0.0002\\
\hline
\end{tabular}
\begin{tablenotes}
\item (a) Line flux ratio on 2020-05-20 were produced for $f=0.5$, while for other dates, $f=0.1$ was used.
\item (b) The root mean square (\textit{rms}) value was computed as a measure to evaluate the overall differences between the observed and generated flux ratio \citep[see][]{2024MNRASHabtie}. This provides an approximation of the mean disparity between the two.
\end{tablenotes}
\end{threeparttable}
\end{table}
\par
In the following section, we attempted to reproduce the coronal emission lines observed simultaneously with the dusty epoch from June 2020 to September 2020 in V2891 Cyg using collisional ionization models of \textsc{cloudy}.


\subsection{Coronal emission lines}\label{sec:coronal}

During the late stages (about 250 days), the spectra of V2891 Cyg exhibited the emergence of coronal emission lines. On June 07, the NIR spectrum showed the presence of many coronal emission lines, predominantly including [\ion{Si}{x}] 1.4305 $\mu$m, [\ion{Si}{vi}] 1.9650 $\mu$m, [\ion{Fe}{xii}] 2.2170 $\mu$m, [\ion{Al}{vi}] 3.6600 $\mu$m, and [\ion{Si}{ix}] 3.9357 $\mu$m. The coronal emission features arising from [\ion{Ti}{vi}] 1.7156 $\mu$m, [\ion{Al}{ix}] 2.0400 $\mu$m, and [\ion{Ca}{viii}] 2.3211 $\mu$m were seen in the spectrum on June 30. These characteristics remained until September 19 (see Figure 15 in \citet{2022MNRAS.510.4265K}). All identified coronal emission lines have an ionisation potential greater than 100 electron volts (eV). The coronal emission lines are classified as forbidden emission lines that originate from the highly ionised state of heavy metals. In the IR spectra of novae, coronal lines are frequently observed, but not all novae go through a coronal phase. Generating these lines requires a very energetic environment, typically found in the coronal region of novae ejecta, which might potentially form from photoionization. On the other hand, coronal lines can also be produced due to collisional ionisation, which may occur when strong internal shocks develop within the ejecta. These shocks may arise from collisions between ejecta that are travelling at different velocities or from the interaction between ejected material and an already-existent stellar wind \citep[][and references therein]{Greenhouse1990ApJ,2012BASI...40..243B,2012BASI...40..213E}.
\par
In the study of V2891 Cyg, \citet{2022MNRAS.510.4265K} demonstrated that the temperature range of around 35000–54000 K for the pseudo-photosphere during the coronal line emission phase was inadequate for the generation of coronal emission lines with ionization potentials above 100 eV. Additionally, the authors could not detect any high-temperature source (with a temperature greater than $10^5$ K) through their UV/X-ray observations with \textit{Swift} mission. These findings support the conclusion that a hot central ionising source is unlikely to be present and suggest that the observed coronal lines are primarily generated through collisional ionization in V2891 Cyg \citet{2022MNRAS.510.4265K}.
\par
The forward shock results from the interaction between the nova ejecta and a dense external shell \citep{Metzger2014MNRAS}. The presence of such a dense external medium in the case of V2891 Cyg has previously been discussed by \citet{2022MNRAS.510.4265K}, who attributed the occurrence and subsequent dissipation of absorption features in the \ion{H}{$\alpha$} line profile to the interaction of the ejecta with the dense external medium. Additionally, \citet{2019ATel13258....1S} observed a narrowing of the \ion{H}{$\alpha$} line from a Full Width at Half Maximum (FWHM) value of 840 km s$^{-1}$ to 300 km s$^{-1}$ between the outburst and November 2019, which they associated with the deceleration of the ejecta by a dense surrounding medium. Based on the theoretical framework proposed by \citet{Metzger2014MNRAS}, we adopted the idea that a highly energetic environment is formed at the outer region of the ejecta, which is heated by the forward shock. The phenomenology is shown in Figure~\ref{fig:model}. We considered a very thin spherical shell outside the ejecta at a distance given by $R_{out} + \Delta_s$, where $\Delta_s = 10^{10.80}$ cm. To reproduce these observed flux ratios, we varied the temperature and density of this region and chemical abundances in small steps. We selectively varied the elements that exhibited coronal emission lines in the spectra while maintaining the remaining elements at their solar values, as given by \citet{2010Ap&SS.328..179G}. We incorporated all the available energy levels from the \textsc{chianti} database \citep{2012ApJLandi} in our \textsc{cloudy} models to reproduce these observed coronal emission lines, as these levels are not part of the default atomic and molecular \textsc{stout} database \citep{2015ApJLykins} in \textsc{cloudy}.
\par
As discussed by \citet{2022MNRAS.510.4265K} that photoionization from the central ionising source didn't play any role in generating these observed coronal lines, we simulated the \textsc{cloudy} models by considering only collisional ionization. We found that our models have successfully generated the observed flux ratios in such cases. We also tried to reproduce the line flux ratio using photoionization as a test case, with the source parameter as estimated from the best-fit model discussed in section~\ref{sec:centralsource}; however, such exercises failed to produce the observed flux ratio of coronal emission lines. The generation of coronal lines is very sensitive to the temperature parameter of the models. We observed that [\ion{Si}{x}], [\ion{Fe}{xii}], and [\ion{Si}{ix}] did not appear at low temperatures ($< 10^{5.5}$K). [\ion{Si}{ix}] appeared at around $\sim 10^{5.6}$ K, while the emission line [\ion{Si}{x}] emerged at around $\sim 10^{5.65}$ K. The emission line [\ion{Fe}{xii}] appeared after $10^{5.7}$ K. Our analysis has shown that these coronal emission lines are best generated at the temperature around $\sim 7.49 \times 10^5$ K, $\sim 7.16 \times 10^5$ K, and $\sim 7.07 \times 10^5$ K for the June 07, June 30, and September 19 dates, respectively. The density of the coronal region was estimated to be around $4.36 \times 10^8$ cm$^{-3}$ on June 07 and June 30, which further increased to $1.58 \times 10^9$ cm$^{-3}$ for Sept 19 (see Table~\ref{tab:coronallines}). These simulations and derived parameters, thus, substantiate the arguments presented by \citet{2022MNRAS.510.4265K} that the observed coronal emission features in V2891 Cyg are attributed to collisional ionization rather than photoionization from the central ionizing source. A highly energetic environment has been established as a result of shock heating, which is accountable for the production of these infrared coronal emission lines of IP $> 100$ eV in nova V2891 Cyg. \citet{Greenhouse1990ApJ} conducted a study on the coronal emissions seen in three novae, namely Nova V1819 Cyg, Nova V827 Her, and Nova QU Vul. Their investigation led to the estimation that the temperature of the coronal regions inside these novae is around $3 \times 10^5$ K. \citet{Chandrashekar1993JApA} estimated the density of the coronal regions to be $2 \times 10^9$ cm$^{-3}$ by assuming that the shock-heated line emitting region is present at the outer periphery of the dust-forming region in Nova Herculis 1991. Our present studies show similar physical conditions are responsible for producing coronal emission in V2891 Cyg.


\begin{table*}
\caption{NIR coronal emission lines flux w.r.t the [Si VI] 1.9650 $\mu$m in V2891 Cyg.\label{tab:coronallines}}
\setlength{\tabcolsep}{5pt}
\small
\begin{threeparttable}
\centering
\begin{tabular}{l c c c c c c c ccccc}
\hline
\hline
Species &  Wavelength &  \multicolumn{2}{c}{June 07} & \multicolumn{2}{c}{June 30} & \multicolumn{2}{c}{Sep 19} & Remarks\\
        &   ($\mu$m)  &  Observed & Modeled$^{a}$ &  Observed & Modeled$^{b}$ &  Observed & Modeled$^{c}$  \\
        &             &  flux & flux         &  flux & flux        &  flux & flux   \\        
\hline
$[$Si X$]$ &  1.4305  &  $3.50 \times 10^{-1}$ & $3.51 \times 10^{-1}$ & $1.26 \times 10^{-1}$ & $1.30 \times 10^{-1}$ & ... & ... & ...\\
$[$Ti VI$]$ & 1.7156  & ... & ... & $4.86 \times 10^{-3}$ & $4.87 \times 10^{-3}$ & ... & ... & ...\\
$[$Si VI $]$ & 1.9650  & 1.00 & 1.00 & 1.00 & 1.00 & 1.00 & 1.00 & Deblended from Br 8-4\\
$[$Al IX$]$ & 2.0400 &  ... & ...& $3.64 \times 10^{-2}$ & $1.09 \times 10^{-1}$ & $1.44 \times 10^{-2}$ & $5.10 \times 10^{-1}$ & ...\\
$[$Fe XII$]$ & 2.2170 &  $9.22 \times 10^{-2}$ & $9.20 \times 10^{-2}$ & $2.43 \times 10^{-2}$ & $2.43 \times 10^{-2}$ & $1.24 \times 10^{-2}$ & $1.41 \times 10^{-2}$ & ...\\
$[$Ca VIII$]$ & 2.3211 & ... & ... & $1.00 \times 10^{-1}$ & $1.09 \times 10^{-1}$ & $5.57 \times 10^{-2}$ & $5.99 \times 10^{-2}$& ...\\
$[$Al VI$]$ & 3.6600 &  $2.71 \times 10^{-1}$ & $2.75 \times 10^{-1}$ & ... & ... & ... & ... & Deblended from Hu 19-6\\
$[$Si IX$]$ & 3.9357 & $1.78 \times 10^{-2}$  & $1.92 \times 10^{-2}$ & ... & ... & ... &... & ...\\
\hline
\end{tabular}
\begin{tablenotes}
\item (a) Generated by considering T$_{cor} \sim 7.49 \times 10^5$ K, $n \sim 4.36 \times 10^8$ cm$^{-3}$, Al/Si $\sim$ 17.6, Fe/Si $\sim$ 28.3. 
\item (b) T$_{cor} \sim 7.16 \times 10^5$ K, $n \sim 4.36 \times 10^8$ cm$^{-3}$, Fe/Si $\sim$ 21.6, Ca/Si $\sim$ 4.16, Ti/Si $\sim$ 60. 
\item (c) T$_{cor} \sim 7.07 \times 10^5$ K, $n \sim 1.58 \times 10^9$ cm$^{-3}$, Al/Si $\sim$ 10.8, Fe/Si $\sim$ 18.33, Ca/Si $\sim$ 3.83, Ti/Si $\sim$ 51.6 . 
\end{tablenotes}
\end{threeparttable}
\end{table*}

\section{Summary and Conclusion}\label{sec:summary}

In this study, we investigated classical nova V2891 Cyg, which erupted in 2019. V2891 Cyg showed some intriguing characteristics, such as time-varying P-Cygni profile only in \ion{O}{i} 7773 {\AA}, a brief dust formation epoch coinciding with coronal emission, and rapid evolution of \ion{O}{i} 1.3164 $\mu$m line flux during this phase. This nova has provided, most likely, rare observational evidence of shock-induced dust formation in CNe \citep{2022MNRAS.510.4265K}. Recently, another study by \citet{2023ApJBanerjee} also offers rare observations of a recurrent nova where dust and CO formation are likely triggered by strong internal shocks within the ejecta. It has now been well established by various studies that internal shocks do indeed form in novae \citep[][and references therein]{Li2017NatAs,Metzger2016MNRAS,Chomiuk2021ARA&A}. In this study, we employed \textsc{cloudy} photoionization models to estimate various physical and chemical characteristics during the nebular phase. Furthermore, we utilised \textsc{cloudy} collisional ionization models to replicate the emission characteristics of certain lines, such as \ion{O}{i} 1.3164 $\mu$m and coronal emission lines.
\par
Our \textsc{cloudy} best-fit photoionization models successfully reproduced the majority of observational features in V2891 Cyg using a two-component model except for \ion{O}{i} 7773 {\AA} emission line. Subsequently, we introduced a third component—a dense, low-mass component to reproduce the \ion{O}{i} 7773 {\AA} emission line. While our photoionization \textsc{cloudy} models failed to reproduce the observed flux ratio of \ion{O}{i} 1.3164 $\mu$m / Pa $\beta$, collisionally ionized \textsc{cloudy} models effectively generated the observed flux ratios. The emission was effectively reproduced by considering a thin, dense shell, with a density ranging from approximately $10^{12.5}$ to $10^{12.8}$ cm$^{-3}$. This indicates that collisional ionization likely plays a significant role in producing the emission line of \ion{O}{i} at 1.3164 $\mu$m in V2891 Cyg. Moreover, the observed fluxes of coronal emission lines in V2891 Cyg from June 2020 to September 2020 were reproduced by employing collisional ionization models from \textsc{cloudy} rather than relying on photoionization models. It is important to acknowledge that our phenomenological method may not be perfect due to the inherent limitations in \textsc{cloudy} models. Novae are known for their complex ejecta and morphology \citep[see, e.g.,][and references therein]{2008ApJRupen,Chomiuk2014Natur,2018ApJMason}, and our study employed a simplified approach, assuming a spherical shape, as \textsc{cloudy} is a one-dimensional photoionization code.
\par
Numerous studies have proposed that dust formation in novae can arise from various factors influenced by different physical and chemical characteristics and different geometries \citep[e.g.,][and references therein]{1988MNRAS507Rawlings,1989MNRASRawlings,2004MNRASPontefract,Shore2004A&A,Shore2018A&A,2022ApJPandey}, and we do not wish to disregard these possibilities. Nonetheless, our objective in this study was to investigate whether the path proposed by \citet{Derdzinski2017MNRAS} and \citet{Metzger2014MNRAS} could be applicable in the case of nova V2891 Cyg, as argued in \citet{2022MNRAS.510.4265K}. Our phenomenological findings suggest this could be the case in this particular nova. The results of this study could support the conjecture that internal shocks in V2891 Cyg, resulting from collisions between multiple mass ejections and the matter behind the shock region, have led to the formation of a dense, thin shell characterized by high density ($\sim 10^{12.5}$ - $10^{12.8}$   cm$^{-3}$), inside which dust has formed in V2891 Cyg. These findings align with the theoretical framework developed by \citet{Derdzinski2017MNRAS}. The apparent correlation of drop in \ion{O}{i} 1.3164 line flux, which coincided with the coronal and dust epoch, is an intriguing feature and could be related to the shock-induced dust formation in V2891 Cyg. If such is the case, this \ion{O}{i} line may act as a tracer for dust formation in nova 2891 Cyg.

\section*{Acknowledgments}
We sincerely thank Prof. Aneurin Evans, the referee, for his valuable insights and constructive feedback, which have significantly improved the quality and clarity of our manuscript. The research work at the Physical Research Laboratory (PRL) is funded by the Department of Space, Government of India. RP thanks PRL for her Post-doctorate fellowship. GS acknowledges WOS-A grant from the Department of Science and Technology (SR/WOS-A/PM-2/2021). PRL operates the Mt. Abu observatory with 1.2m and 2.5m telescopes at Mt. Abu. We acknowledge use of data collected from PRL 1.2m telescope 
 at Mt. Abu Observatory with MFOSC-P instrument. We also acknowledge the use of NIR spectra collected from IRTF telescope with SpeX instrument.

\section*{Data Availability}
Simulations in this paper made use of the code \textsc{cloudy} (c22), which can be downloaded freely at \url{https://www.nublado.org/}. The observed data have already been published \citep{2022MNRAS.510.4265K}.

\newpage
\bibliographystyle{mnras}
\bibliography{references} 





\bsp	
\label{lastpage}
\end{document}